\documentclass[traditabstract]{aa}
\usepackage{graphicx}
\usepackage{natbib}
\usepackage{txfonts}

\begin{document}

\title{Turbulence in the ICM from mergers, cool-core sloshing and jets: results from a new multi-scale filtering approach.}

\author{F. Vazza\inst{1,2}, E. Roediger\inst{1} \& M. Br\"{u}ggen\inst{1}}

\offprints{F. Vazza\\ \email{f.vazza@jacobs-university.de}}

\institute{Jacobs University Bremen, Campus Ring 1, 28759, Bremen, Germany
\and  INAF/Istituto di Radioastronomia, via Gobetti 101, I-40129 Bologna, Italy}

\date{Received / Accepted}

\authorrunning{F. Vazza, E. Roediger \& Br\"{u}ggen }
\titlerunning{A multi-scale method to measure turbulence.}

\abstract
{We have  designed a simple multi-scale method that identifies turbulent motions in hydrodynamical grid 
simulations. The method does not assume any a priori coherence scale to distinguish laminar and turbulent flows. Instead, the local mean velocity field around each cell is reconstructed  with a multi-scale filtering technique, yielding the maximum 
scale of turbulent eddies by means of iterations.
The method is robust, fast, and easily applicable to
any grid simulation.
We present here the application of this technique to the study
of spatial and spectral properties of turbulence in the intra-cluster
medium, measuring turbulent diffusion and anisotropy of the turbulent
velocity field for a variety of driving mechanisms:
a) accretion of matter in galaxy clusters (simulated with {\small ENZO}); 
b) sloshing motions around cool-cores (simulated with {\small FLASH}); c) jet outflows from active galactic nuclei, AGN, (simulated with {\small FLASH}).
The turbulent velocities driven by matter accretion in galaxy clusters are mostly tangential in the inner regions (inside the cluster virial radius) and isotropic in regions close to the virial radius. The same is found for turbulence excited by cool-core sloshing, while the jet outflowing from AGN drives mostly radial turbulence motions near its sonic point and beyond. Turbulence leads to a diffusivity in the range $D_{\rm turb} \sim 10^{29}-10^{30} ~{\rm cm^{2} ~s^{-1}}$ in the intra-cluster medium. On average, the energetically dominant mechanism of turbulence driving in the intra cluster
medium is represented by accretion of matter and major mergers during
cluster evolution.}

\maketitle


\section{Introduction}
\label{sec:intro}


On many scales, astrophysical fluids show signs of turbulence whose dynamical contribution may 
range from significant, as in the case of the intra cluster medium (ICM) \citep[e.g.][]{bn99,do05,su06} to dominant, as in the case of the
interstellar medium (ISM)  \citep[e.g.][]{la81,g_s95,padoan02,maclow04}.
Turbulence is a fundamental phenomenon that provides viscosity in accretion disks \citep[e.g.][]{brandenburg95,balbus98}, that transports matter in stellar atmospheres \citep[e.g.][]{canuto91} and mixes high- and low-metallicity ICM in cluster cores \citep[e.g.][]{re06}.

The direct numerical simulations of turbulence need to follow the turbulent
cascade over a wide range of length scales. Recently, this has become feasible, as hydrodynamical simulations can reach fairly wide dynamic ranges
(of $\sim 2-3$ orders of magnitude in scales, e.g. \citealt{jones11} and references therein). However, these high-resolution simulations do not resolve
the length scale of physical turbulent dissipation, and subgrid turbulent
closures that incorporate the evolution and effect of turbulence
on unresolved scales have been developed \citep[e.g.][]{schmidt06,sb08,maier09}.
 
 \bigskip
 
The analysis of turbulence simulations of realistic systems (e.g. galaxy
clusters) requires the separation of bulk and turbulent flows, and a number
of strategies have been proposed in the recent past.
A simple method would be that of computing the turbulent velocity field
as the residual respect to the ICM velocity field, averaged over spherical
shells from the cluster centre \citep[][]{bn99,in08,lau09}. Alternatively,
on can compute the average velocity field of the ICM via 3--D interpolation,
and consider as turbulent the velocity structure below the interpolation
scale \citep[][]{do05,va06,va09turbo,va11turbo,valda11}.  Alternative
approaches focus on the decomposition of solenoidal and rotational
components of the velocity field \citep[][]{ryu08,zhu10}, or employ
sub-grid modelling \citep[][]{maier09,iap11}.
These methods a priori assume limiting length scales of turbulence,
possibly leading to inconsistent results. For instance, 
 for a similar cluster mass the estimated amount of turbulent pressure in the cluster core may range from $\sim 0.2$ percent of the total gas pressure using sub-grid modelling estimates (Maier et al.2009) to $\sim 2$ percent of the total gas pressure by filtering the velocity
field with a  radial average (Iapichino \& Niemeyer 2008), to $\sim 2-5$ percent by using a filtering scale of $\approx 300$ kpc \citep[][]{va09turbo}.

\bigskip

In this article, we propose a method that locally determines the velocity
coherent scale and uses this to distinguish the laminar and the turbulent
components of the velocity field. Thus, the method needs no a priori assumptions of the typical scales of the flow in the simulated volume (Sec.\ref{sec:methods}). The performances of our method are tested 
 with an idealised setup in Sec.\ref{subsec:test}.
We present the first results on the properties of turbulence stirred by
major mergers and accretions (Sec.\ref{subsec:enzo1}), gas sloshing in cool cores (Sec.\ref{subsec:cool_core}), and active galactic nuclei (AGN) outbursts (Sec.\ref{subsec:agn}).
Our conclusions are given in Section \ref{sec:conclusions}; in the
appendix we give an example of our algorithm in {\small IDL 7.0} syntax.

\begin{figure*}
\begin{center}
\includegraphics[width=0.95\textwidth]{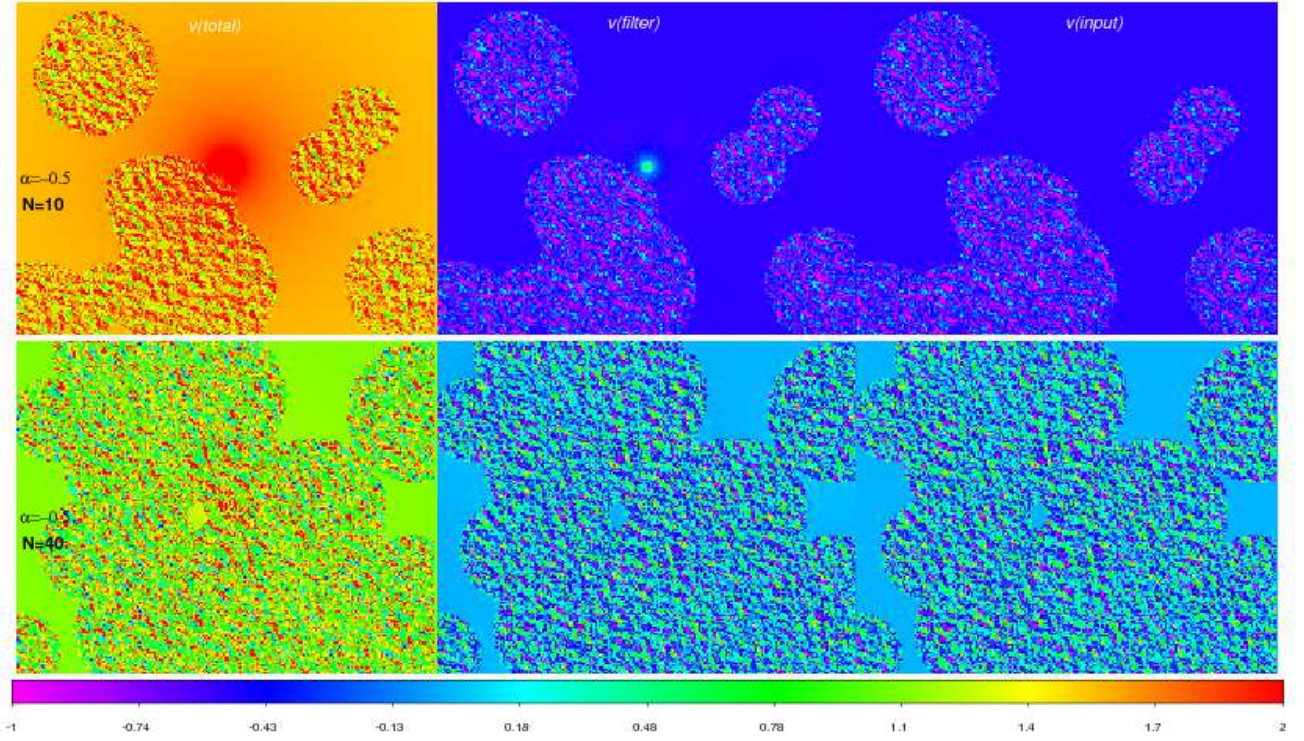}
\caption{Maps of absolute value of the total velocity field (left, in arbitrary code units), of the turbulent velocity field reconstructed with our method (centre) and of the input turbulent velocity field (right) for two tests with a different slope for the background velocity profile ($\alpha$) and for the number of injected
turbulent patches, $N$ (see Sec.\ref{subsec:test} for details). In the top row we assumed $\alpha=-0.5$,  $N=10$ and $\sigma_{\mathrm v}/{\mathrm v_{\rm tot}}=0.3$, while in the bottom row we assumed  $\alpha=-0.5$,  $N=40$ and $\sigma_{\mathrm v}/{\mathrm v_{\rm tot}}=3$.}
\end{center}
\label{fig:test1}
\end{figure*}

\begin{figure}
\includegraphics[width=0.45\textwidth]{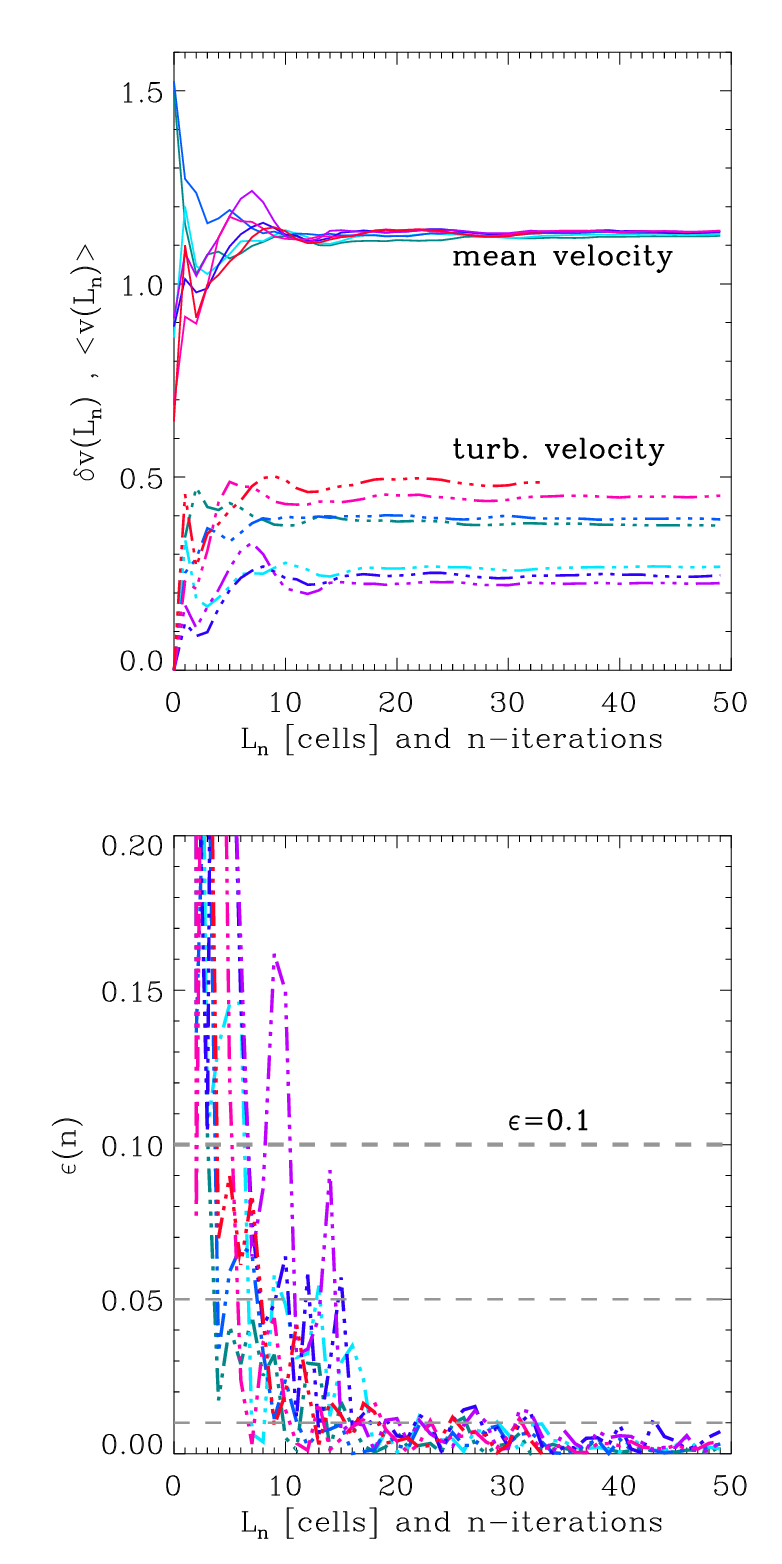}
\caption{Trend with number of iterations of the mean local velocity field and turbulent velocity field reconstructed by our method for eight random points extracted in the first test of Fig.\ref{fig:test1}. The top panel shows how the mean velocity and the turbulent velocity of each points change as the number of iterations is increased (the iterations are stopped when $\epsilon \leq 10^{-3}$; the lower panel  shows the trend of the fractional increase of the turbulent
velocity field with the number of iterations (Eq.\ref{eq:iter}) for the same points. Our fiducial threshold value to stop the iterations, $\epsilon=0.1$, as well as $\epsilon=0.05$ and $\epsilon=0.01$ is
shown for comparison.}
\label{fig:iter}
\end{figure}

\section{Multi-scale filtering}
\label{sec:methods}

Turbulent
fluids are characterised by a hierarchy of scales, ranging from the injection or driving scale, $L_{\rm o}$, down to the 
physical dissipation scale, $\lambda_{\rm diss}$, which sets the minimum scale available to the motion \citep[e.g.][]{landau66,shore07}. 
In incompressible turbulence, the flux of kinetic energy across spatial scales is constant and, if the flow is stationary and uniform, the spectral energy distribution
for scales smaller than $L_{\rm o}$ is described by $E(k) dk \propto k^{-5/3} dk$ \citep[][]{kolmogorov41}. 

This  power-law translates into a simple relation between the physical size
of turbulent eddies, $l$, and their internal velocity dispersion, $\sigma_{\mathrm v}$:

\begin{equation}
\sigma_{\mathrm v}^{2} \sim l^{2/3},
\end{equation}

\noindent (e.g. Landau \& Lifshitz 1966). In most  available numerical schemes, however, the smallest scale available to the fluid motions is much
larger than the physical dissipation one, thus breaking the power-law
behaviour at some numerical scale ($\sim 4-8$ cells in most grid codes, \citealt[e.g.][]{pw94}).

The scale that contains the maximum kinetic energy  is the "integral scale" \citep[e.g.][]{shore07}, and for homogeneous isotropic turbulence it is given by

\begin{equation}
\Lambda_{\rm I} = \frac{\pi}{2 \sigma_{\mathrm v}^{2}} \int^{k_{\rm diss}}_{0} \frac{E(k)}{k} dk,
\label{eq:L}
\end{equation}

\noindent while the largest correlation scale in the fluid, $\Lambda$, is defined by the maximum of $k \cdot E(k)$.

According to this picture, the flow structure is uncorrelated for scales $\gg \Lambda$, and the average velocity within this scale tends to the average fluid velocity.
Based on that, we designed a recursive method to compute the
average value of velocity around a cell for increasingly larger
scales, until numerical convergence is achieved.  
The local mean field computed in this way (averaging for $\leq \Lambda$) can be used to compute the turbulent velocity fluctuations inside this scale.

\bigskip

In detail, our algorithm works as follows:
\begin{itemize}
\item at a given {\it n-}th iteration, the components of the local mean velocity field around each cell are calculated as
\begin{equation}
\label{eq:local}
\overline{\mathrm v(L_{\rm n})} =\frac{\sum_{i}(r<L_{\rm n}){\mathrm v_{\rm i}\cdot w_{\rm i}}} {\sum_{i} w_{\rm i}};
\end{equation}

where $w_{\rm i}$ is a weighting function (e.g. gas density or gas mass);

\item we compute the local "turbulent" velocity field at each  {\it n-}th iteration:
\begin{equation}
\label{eq:deltav}
\delta \mathrm v(L_{\rm n})={\mathrm v}-{\overline{\mathrm v(L_{\rm n})}};
\end{equation}

\item the local and the turbulent velocity field are computed at each
{\it n}-th iteration for increasing values of $L_{\rm n}$  until the
relative variation of turbulent local velocity between two iterations
is below the given tolerance parameter, $\epsilon$:

\begin{equation}
\label{eq:iter}
\frac{\delta \mathrm v(L_{\rm n})-\delta \mathrm v(L_{\rm n-1})}{\delta \mathrm v(L_{\rm n-1})} \leq \epsilon.
\end{equation}
 
In our  case, $\Delta L=L_{\rm n}-L_{\rm n-1}$ is bound to be the minimum available cell size. The parameter $\epsilon$ is a small tolerance parameter, whose value
is tuned by testing (we usually adopt $\epsilon \leq 0.1$, see below).

\item Once convergence of Eq.~\ref{eq:iter} is reached, we fix $\Lambda=L_{\rm n}$ and ${\overline{\mathrm v(L_{\rm n})}}={\overline{\mathrm v_{\rm \Lambda}}}$, 
and we compute the local turbulent velocity field of the cell as

\begin{equation} 
\label{eq:turbo_vel}
 {\delta \mathrm v}={\mathrm v}-{\overline{\mathrm v_{\rm \Lambda}}}.
\end{equation}
\end{itemize}

This procedure is repeated separately for each velocity component.
In simulations using the piecewise parabolic method (PPM) we set a minimum
radius of $R_{s}=4 \Delta x$ at the start of iterations, since smaller
scales can be affected by numerical dissipation.

Note that if we choose the gas density as the weighting function, $w_{\rm i}$, in Eq.~\ref{eq:local} the numerical noise potentially arising by having gas cells at lower resolution far away from the cell location is minimised. This makes the scheme also readily applicable to 
SPH, after interpolation onto a regular mesh.

\bigskip
The numerical noise produced near strong shocks in the
simulated volume affects the correct measurement of $\Lambda$ and
$\vec{\overline{\mathrm v_{\rm \Lambda}}}$. In this case, the convergence of Eq.\ref{eq:iter} is made slower because two different pre-shock and post-shock velocities are averaged across the shock. Strong shocks are characterised by 
a highly skewed distribution of velocities across the shock, and therefore monitoring the skewness of
each velocity component in the volume around each cell is an
efficient way of identifying the contribution from shocks to the 
local estimate of ${\overline{\mathrm v_{\rm \Lambda}}}$. 
In detail, prior to our analysis we measure
the skewness of the velocity field (separately for each
component) in volumes of $N_{\rm S}=8^{3}$ cells around each cell in the
domain

\begin{equation}
S_{\rm i}=\frac{1}{N_{\rm S}-1}\sum_{i=1}^{N_{\rm S}}\frac{({\mathrm v_{\rm i}}-{\overline{\mathrm v_{\Lambda}}})^{3}}{(\sigma_{\mathrm v,i})^{3}};
\end{equation}

\noindent where $\sigma_{\mathrm v,i}$ is the variance of the velocity component
inside the $N_{\rm S}$ volume around each cell.
At each iteration of Eq.~\ref{eq:iter}, we measure an average skewness
inside the radius of integration by volume-averaging 
the previously measured values of $S_{\rm i}$, $\overline{S_{\rm n}}=\sum_{\rm i}{S_{\rm i}}/N_{\rm cell}(L_{\rm n})$, where $N_{\rm cell}(L_{\rm n})$ is the number of cells within the integration radius, $L_{\rm n}$. If
a shock enters the integration volume, the average skewness around the cell becomes 
rapidly very large, and the iterations are stopped to avoid
strong contaminations from velocity jumps at shocks. 
We found that stopping the iterations when
$\overline{S_{\rm n}} \geq \epsilon_{\rm S}$, with $\epsilon_{\rm S}=1$ provides reliable results for our
simulated ICM with PPM methods. Using as a reference the realistic case of high-resolution {\it {\small ENZO}} simulated galaxy
clusters (as in Sec.\ref{subsec:enzo1}), we verified that 
slighly different choices 
in the range $\epsilon_{\rm S}=0.5-3$, or in the number of 
cells used for the local estimate of skewness, $N_{\rm S}=5^{3}-15^{3}$,  yield very similar
results (with differences at the $\sim$ percent level on the values of turbulent energy) in the final 3--D distributions
of turbulent velocity field with this method.

In principle,  more complex and accurate shock-detecting schemes can be used \citep[e.g.][and references therein]{va11comparison}, usually employing other physical quantities, e.g. gas temperature, pressure, sound speed.
However, in our algorithm we ideally aim at reconstructing the turbulent field using only the geometrical information
on the velocity field. This makes  
the application of our method to a variety of simulations straightforward,  
without requiring fine tuning of parameters.
The only internal parameters needed to stop the iterations of our algorithm are set after our preliminary testing:  one for the convergence of the mean local velocity field ($\epsilon=0.1$) and a second one to remove the spurious contribution of shock waves ($\epsilon_{\rm k}=1$). In a nearly homogeneous gas density distribution, the weighting function in
Eq.~\ref{eq:local} can also be omitted, and therefore the only physical field needed is the velocity field.

Since our algorithm tries to reconstruct the typical scale of the signal in each point in space, this method
is conceptually similar to the wavelet decomposition analysis used in turbulence studies \citep[e.g.][]{muzy91}.
However, in our approach we do not aim at decomposing the 3D flows in its  spatial components, as in the multi-resolution analysis \citep[e.g.][]{mallat89}, but uniquely to constrain
the largest outer scale of turbulence around each cell.

In the appendix, we reproduce the source code of the basic version of the multi-scale filtering technique, written in {\small IDL} syntax for 3D 
distributions.

\begin{figure*}
\begin{center}
\includegraphics[width=0.24\textwidth]{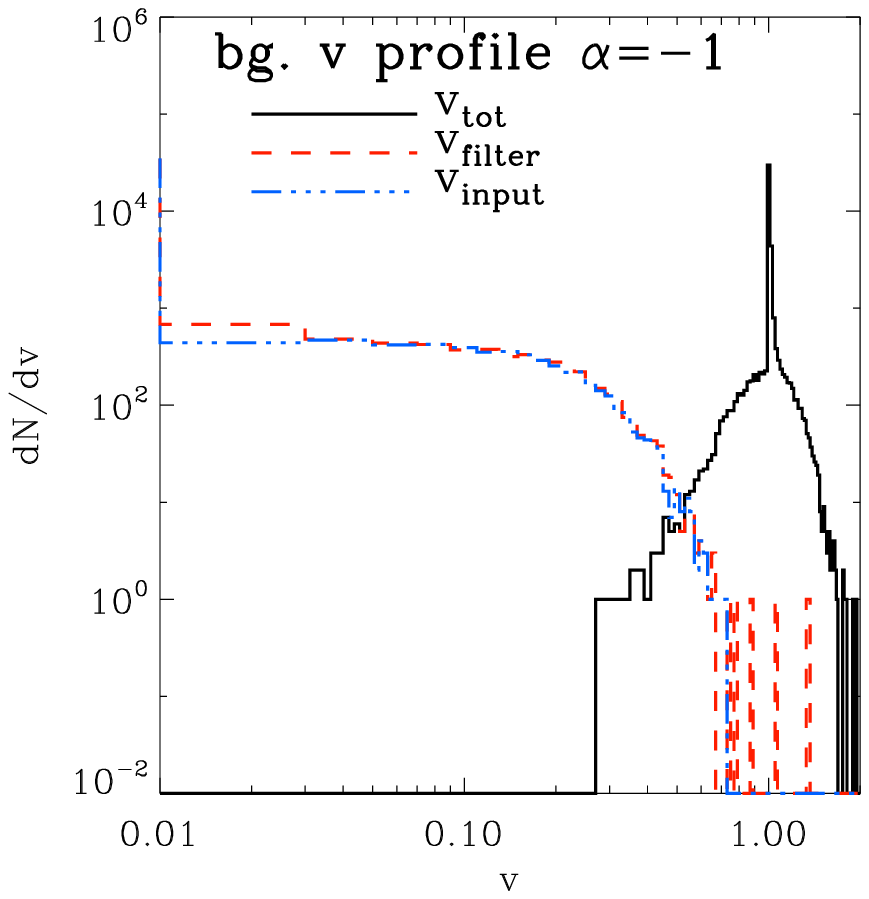}
\includegraphics[width=0.24\textwidth]{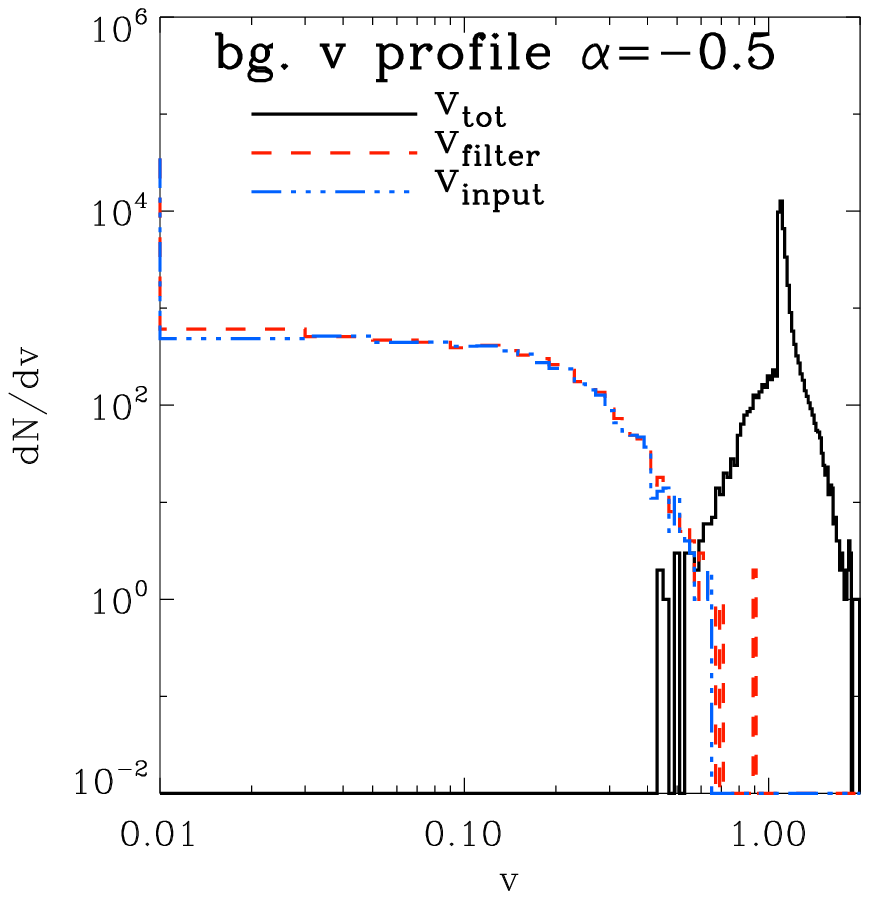}
\includegraphics[width=0.24\textwidth]{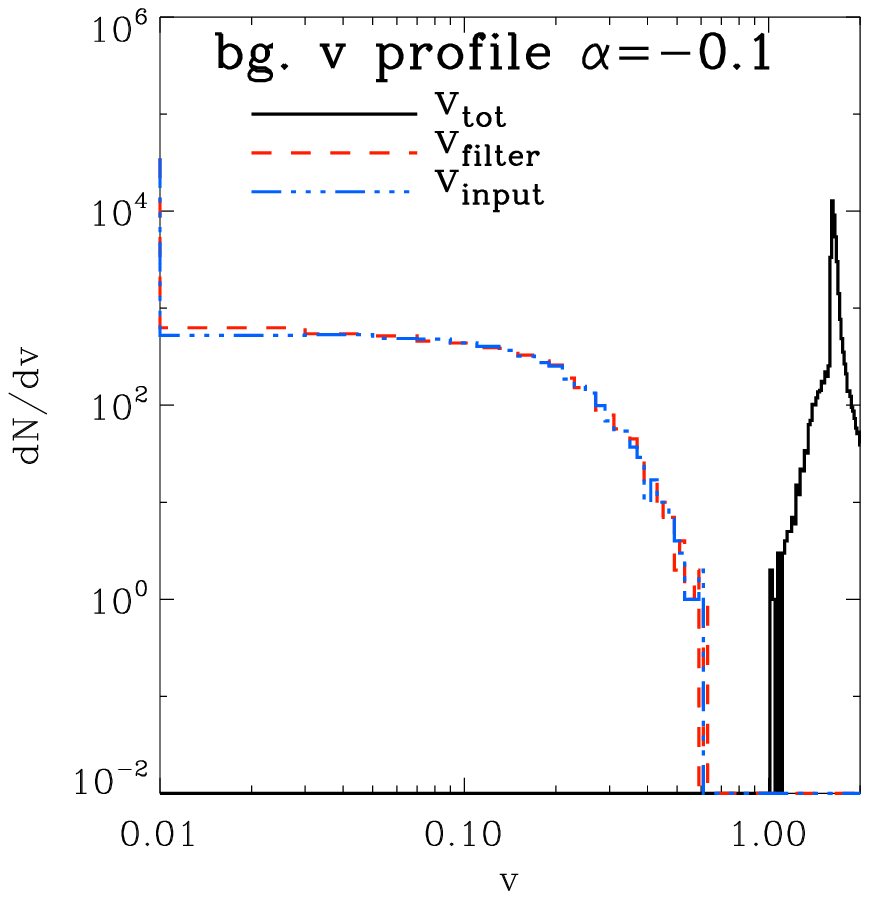}
\includegraphics[width=0.24\textwidth]{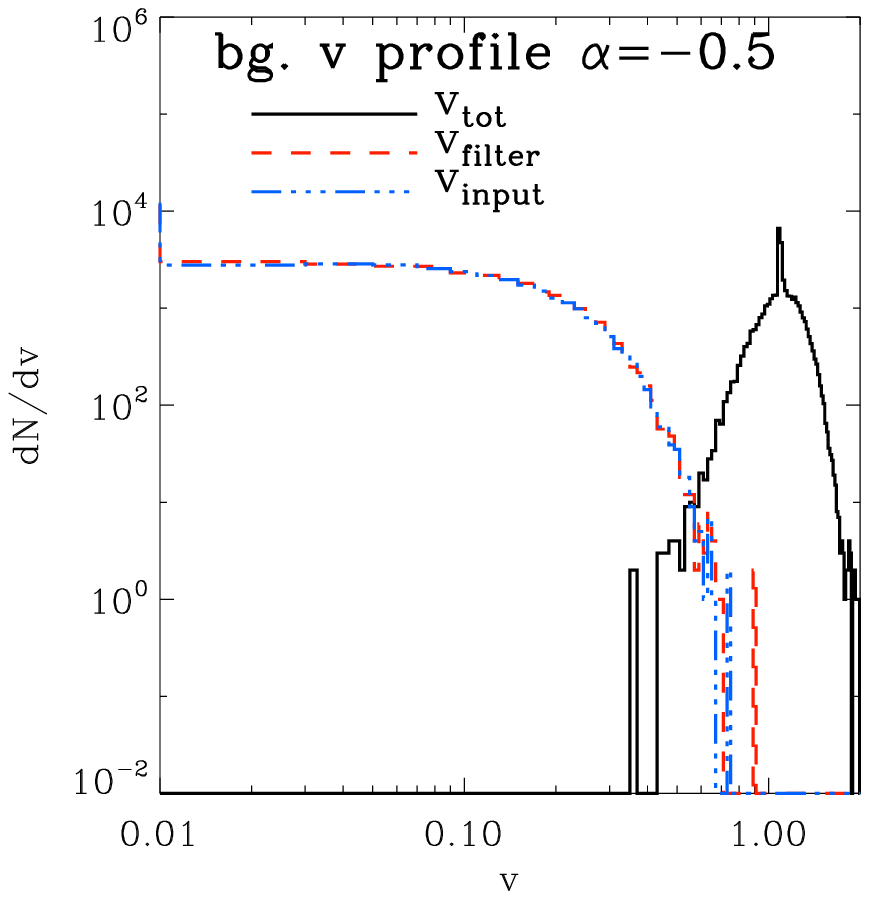}
\includegraphics[width=0.24\textwidth]{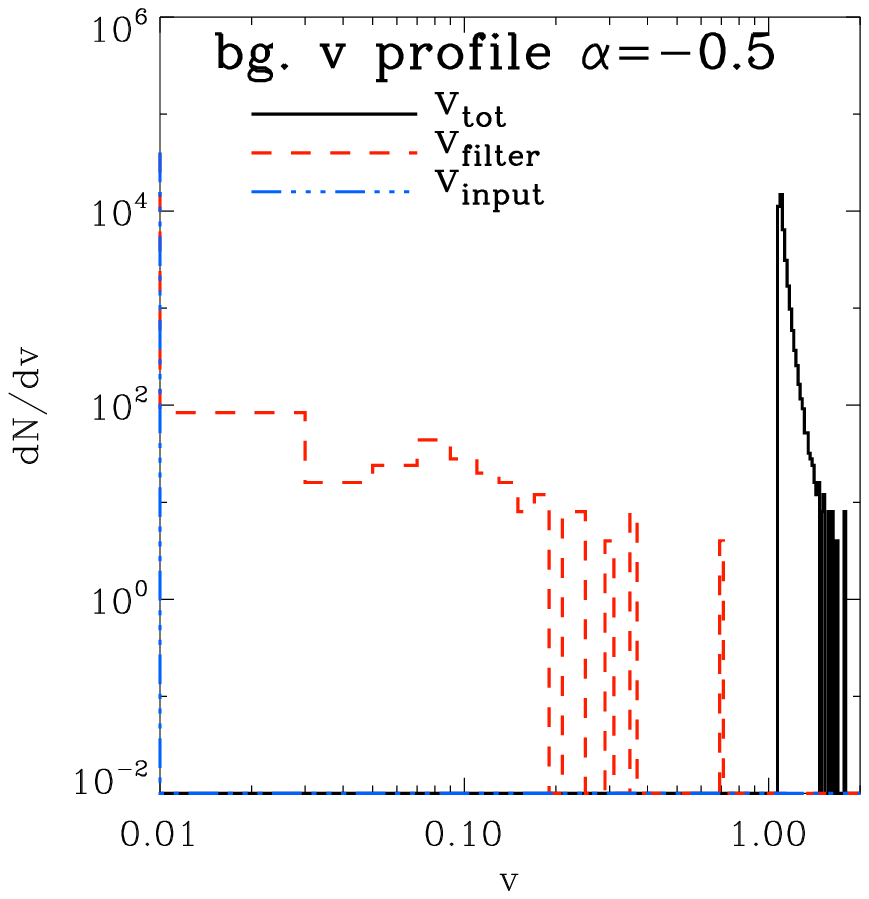}
\includegraphics[width=0.24\textwidth]{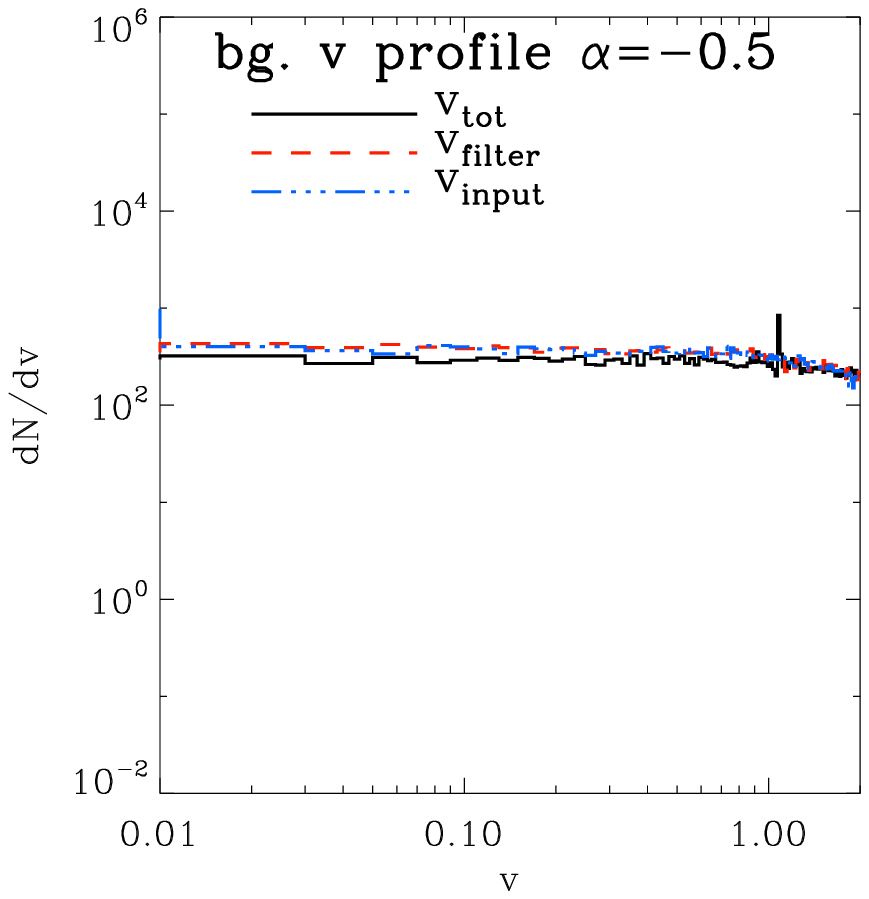}
\includegraphics[width=0.24\textwidth]{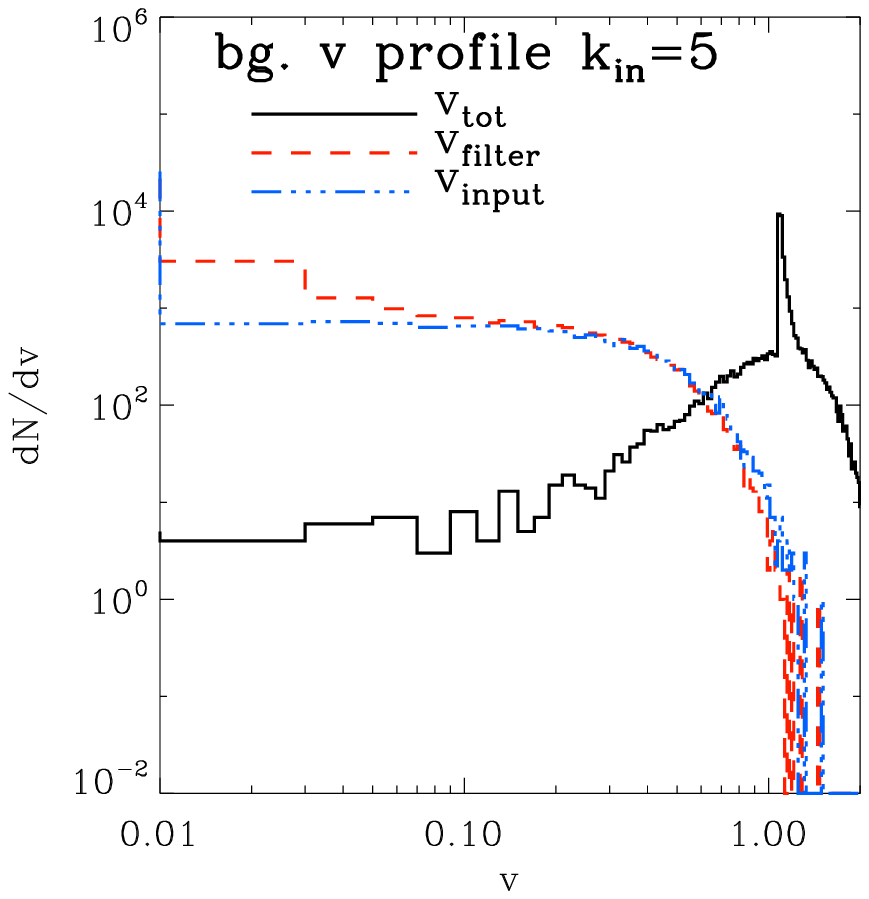}
\includegraphics[width=0.24\textwidth]{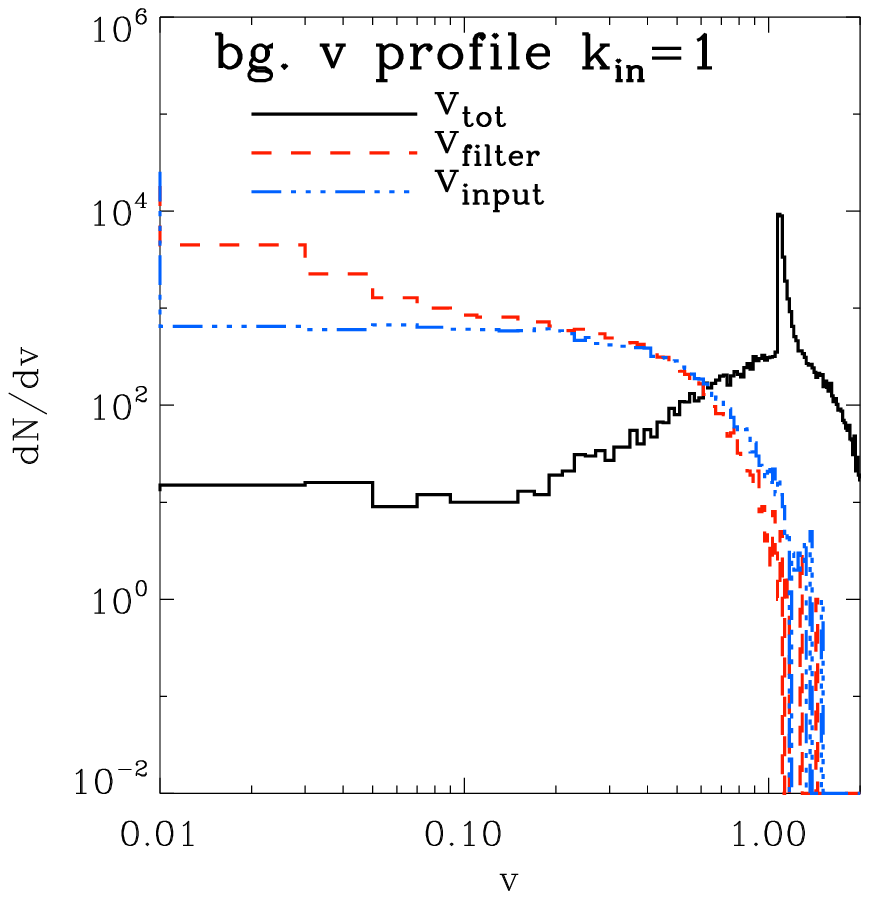}
\caption{Distribution of the velocity modulus for the numerical
tests described in Sec.\ref{subsec:test}. From top left to bottom right:
a) background velocity profile with $\alpha=-1$ and $N=10$ additional patches of turbulent field with $\sigma_{\mathrm v}/\mathrm v_{\rm o}=0.3$; b) same as a), but with $\alpha=-0.5$; c) same as a), but with $\alpha=-0.1$; d) same as b), but with $N=50$;
e) only background velocity field with $\alpha=-0.5$, no turbulent
patches; f) as in b), but with turbulent patches everywhere; g) as in b), but assuming an outer scale of turbulent motions  $k_{\rm in}=5$; h) as in b), but assuming an outer scale of turbulent motions $k_{\rm in}=1$. In each panel, the black lines show the total input velocity field, the blue lines the input turbulent field, and the red lines the 
field reconstructed with our algorithm.}
\label{fig:test2}
\end{center}
\end{figure*}

\begin{figure*}
\begin{center}
\includegraphics[width=0.3\textwidth]{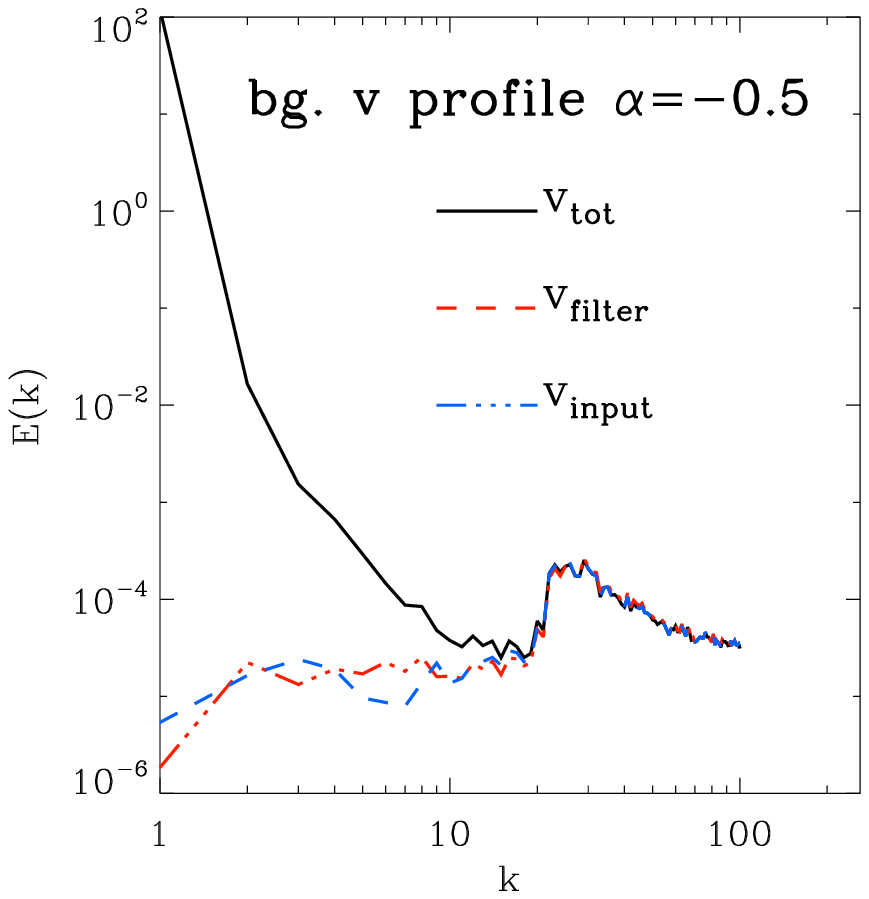}
\includegraphics[width=0.3\textwidth]{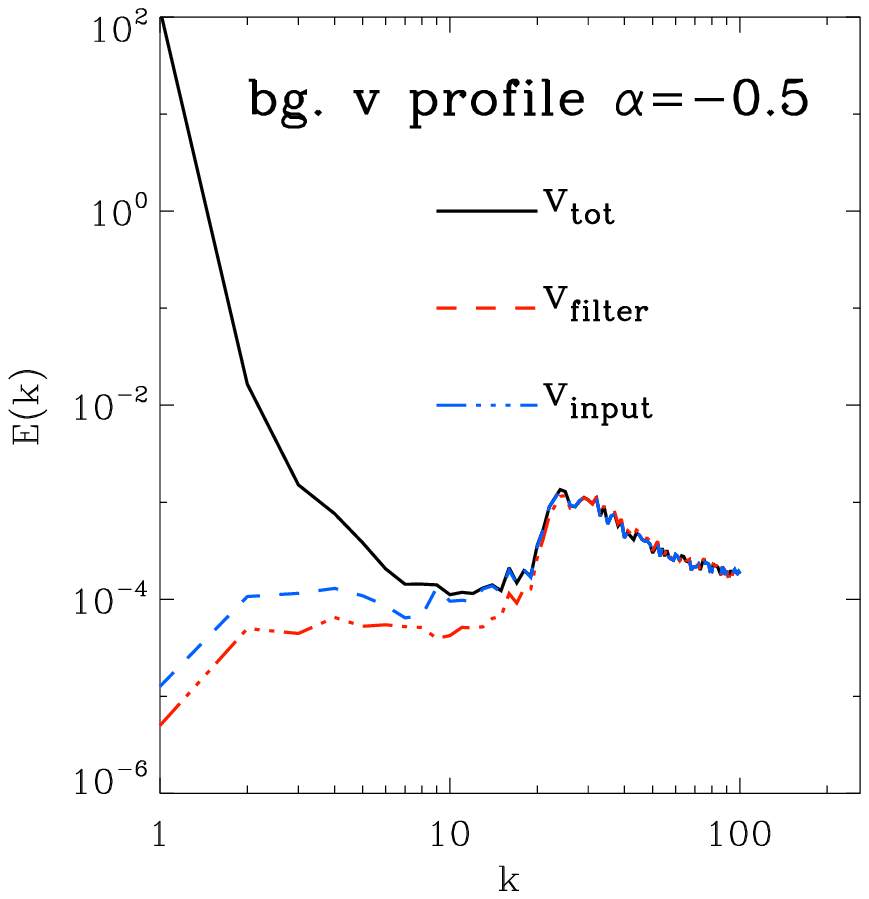}
\includegraphics[width=0.3\textwidth]{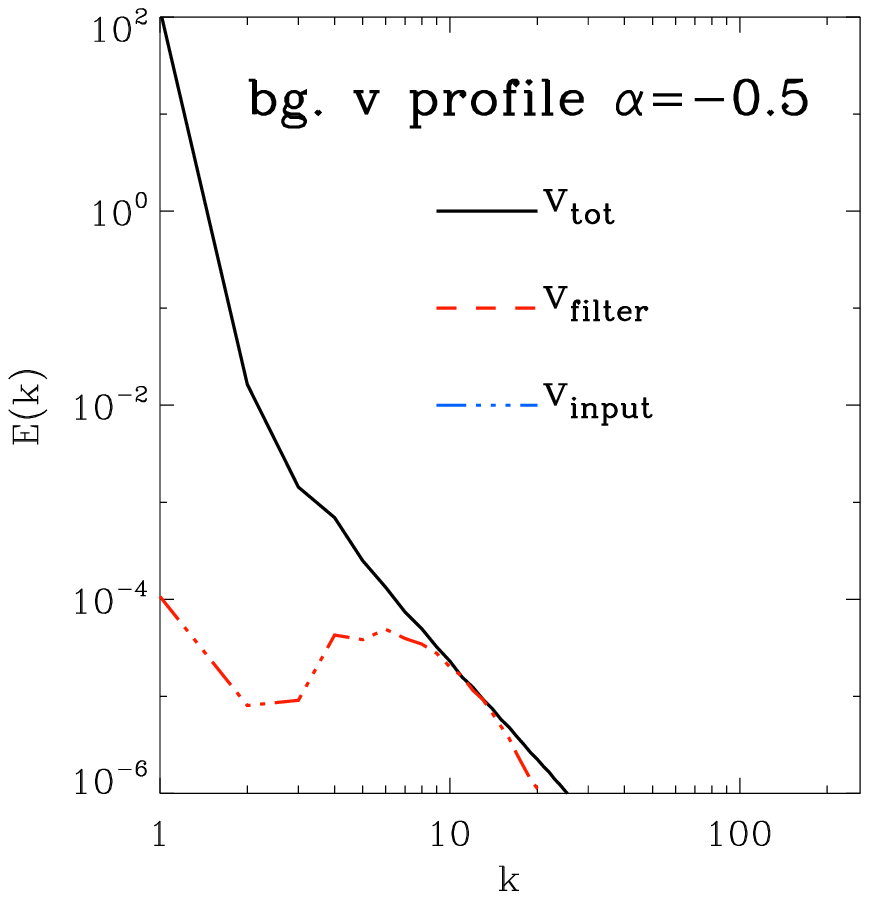}
\includegraphics[width=0.3\textwidth]{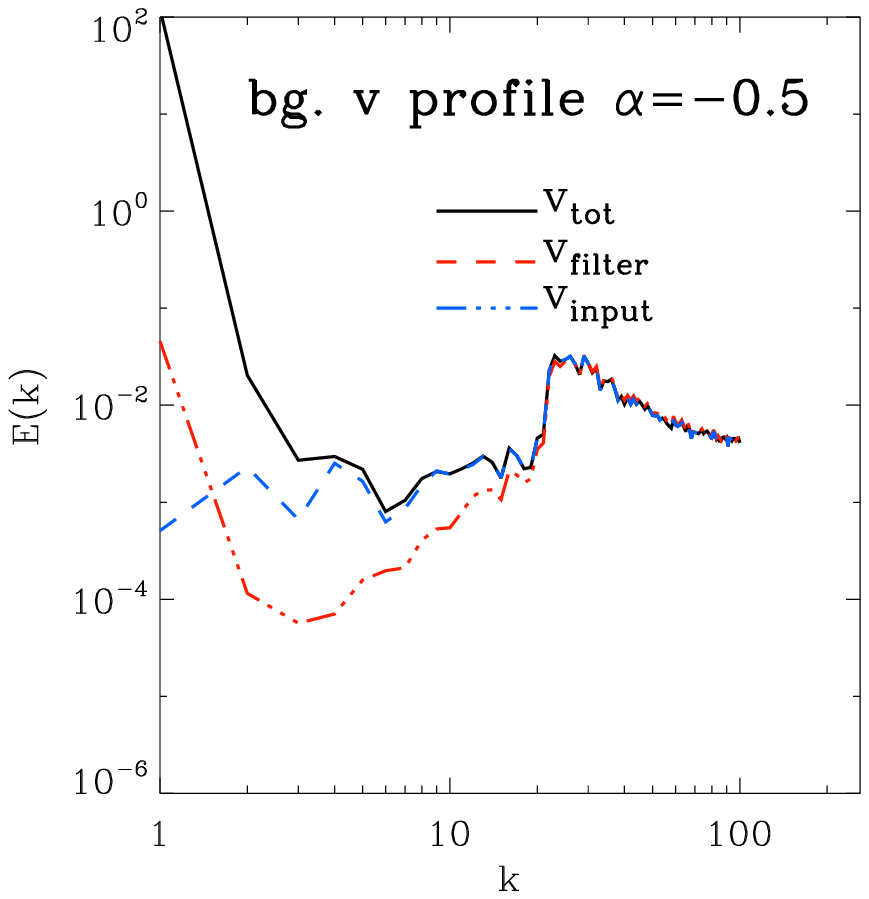}
\includegraphics[width=0.3\textwidth]{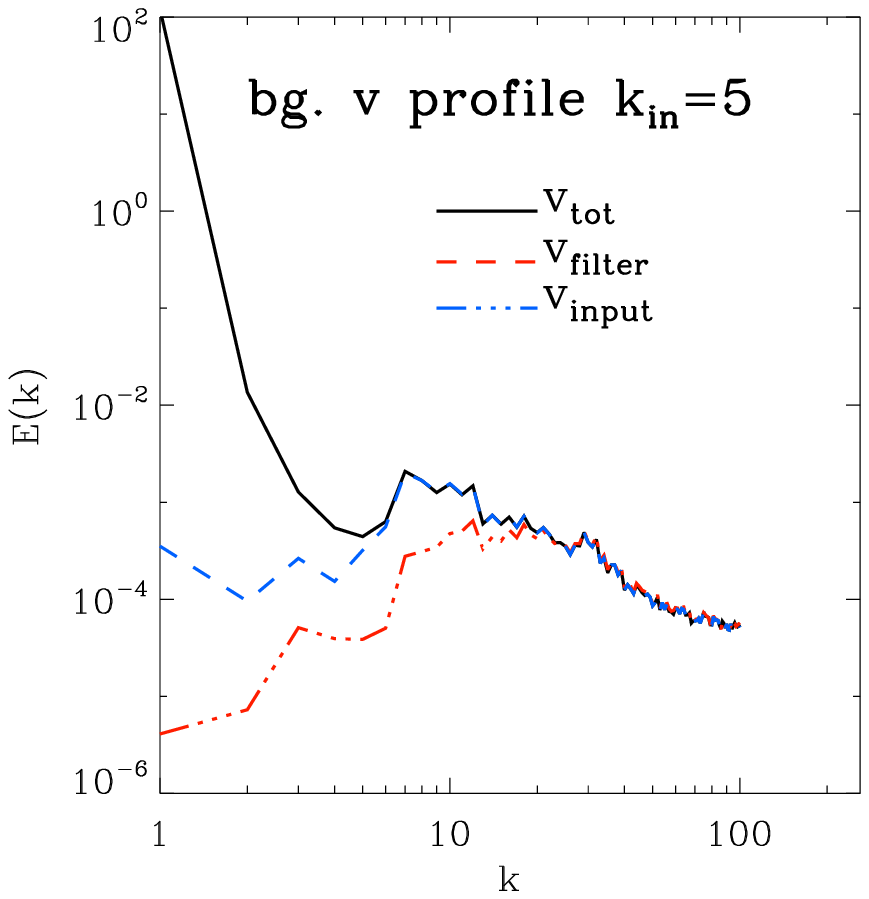}
\includegraphics[width=0.3\textwidth]{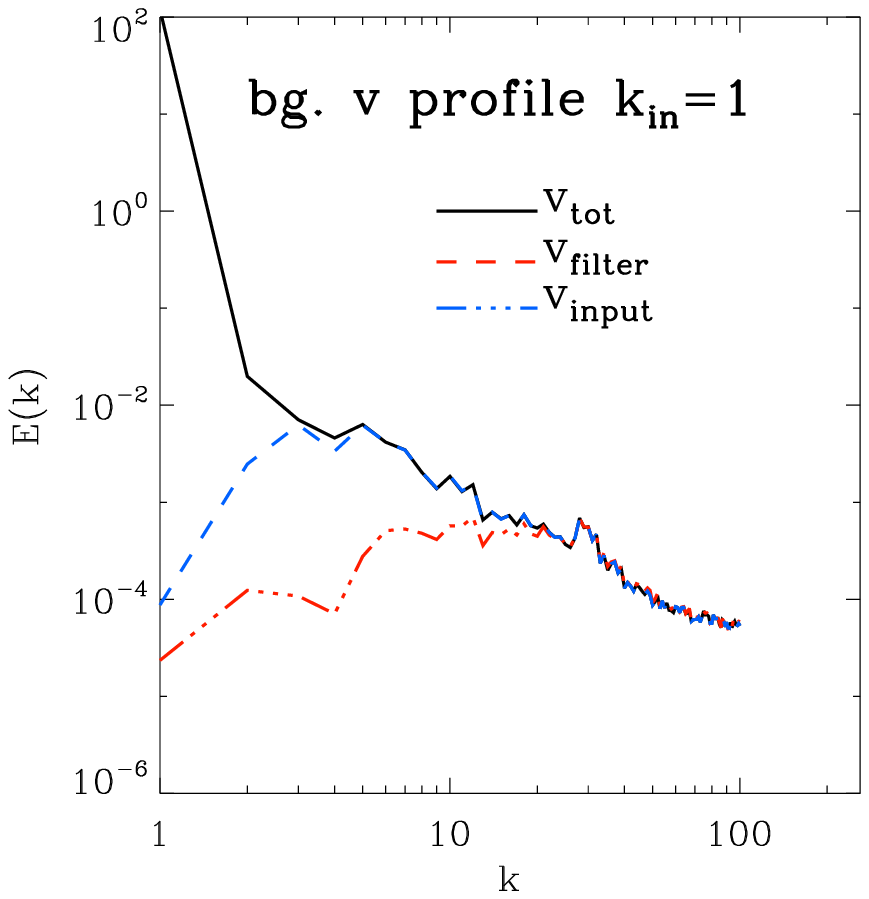}
\caption{Power spectra of velocity field for tests b), d) e)
and f), g) and h) of Fig.\ref{fig:test2}. The black lines show the total input velocity field, the blue lines the input turbulent field, and the red lines the 
field reconstructed with our algorithm. Note that the low $k$ cut-off in the turbulent spectra of test h) is caused by our procedure of extracting small patches of turbulent motions in the simulated volume, while the original turbulent velocity field has by construction an unbroken power-law spectrum from $k=1$ on.}
\label{fig:test3}
\end{center}
\end{figure*}

\subsection{Tests in two dimensions}
\label{subsec:test}
We tested our procedure and our choice of convergence parameters against
idealized 2D setups, in which we constructed combinations of average background velocity field and patches
of chaotic turbulent velocity fields for a $200^{2}$ grid. 
For the regular large-scale velocity field, we set up a
simple radial inflow, according to 
$\mathrm v_{\rm R}(r)=A+B \cdot r^{\alpha}$, where we set $A=B=1$ (in arbitrary code units). We
tested $\alpha=-0.1$, $\alpha=-0.5$ and $\alpha=-1$. These profiles
represent a very generalised version of the profiles of radial velocity found in simulations of the ICM \citep[][]{bn99,fa05}.
We added patches of chaotic velocity field by generating an additional 2D velocity field, $\vec{\mathrm v_{\rm input}}$, with a random extraction from an energy spectrum obeying the $E(k) \propto k^{-5/3}$ law. 
In our fiducial model, we imposed a minimum wavenumber $k_{\rm in}=10$ for the turbulent velocity; we also tested
the cases $k_{\rm in}=1$ and $k_{\rm in}=5$. 
Then we randomly selected $N$ circular regions
with random centres and radii. For the 
areas inside the extracted circular regions, we added the
turbulent field to the background field, $\vec{\mathrm v_{\rm tot}}=\vec{\mathrm v_{\rm o}}+\vec{\mathrm v_{\rm input}}$. 

As an example, we show in Fig.\ref{fig:test1} the maps of the total velocity field created in this way (left), of the 
turbulent velocity field reconstructed with our algorithm (centre) and of the input turbulent of velocity field (right) for two of our tests. 
Figure \ref{fig:iter} shows the convergence of the local mean  velocity and the turbulent velocity with iteration time steps for a random cells extracted in the first test of Fig.\ref{fig:test1}.
Our algorithm on average requires
$5-10$ iterations to converge on $\vec{\delta \mathrm v}$ for each cell, within the $\epsilon=0.1$ tolerance. The second panel of Fig.\ref{fig:iter} shows the behaviour of the fractional change of the
turbulent velocity fluctuation (Eq.\ref{eq:iter}) as a function of
the number of iterations. In the vast majority of cases, stopping the iterations when this fractional change is below $\epsilon=0.1$ represents a very good approximation to constrain the turbulent field
around the cell location. If we let the iterations proceed until a 
fractional variation less than $\epsilon=0.01$ is reached, the number
of iterations increases but the final improvement on the turbulent
velocity field is not substantial. 
Also based on our tests in the more realistic case presented in Sec.\ref{subsec:enzo1}, we suggest that our fiducial choice of $\epsilon=0.1$ is the best compromise between a robust reconstruction of the turbulent field and the speed of the algorithm.

In Fig.\ref{fig:test2} we show the distribution 
of velocities for several tests, comparing the results of our
algorithm to the input laminar and turbulent velocity fields. 

Our method performs well in reconstructing the  original distribution of  turbulent velocities in most cases, with no strong dependence on the background radial profiles (tests a)-c)). Misidentification can happen when the turbulent patches are so numerous that they frequently "merge" into a larger pattern, as in test d). In this case, the algorithm
requires more iterations and a larger scan region to
converge, and the estimated local mean field can be biased in this case. The best morphological reconstruction of turbulent patches is obtained when the turbulent
structures are well separated, and their typical internal
velocity is significantly different from the local mean velocity. 

Sharp cusps in steep
velocity profiles can be misidentified as a turbulent fluctuation,
as shown in test e), where no additional
turbulent field, $\vec{\mathrm v_{\rm input}}$, is added to the
regular radial profile. However, such sharp peaks are unlikely
in realistic simulations of the ICM. Our tests show that the use of $\epsilon=0.1$ is generally the best compromise between the need of a fast convergence of Eq.~\ref{eq:iter} outside of turbulent structures, and the necessity
that the algorithm must not misidentify regular large-scale
gradients as turbulent fluctuations.

A second limitation of our method is that it relies on the assumption 
that the typical scale length of the laminar flow is larger than the
maximum size of turbulent "eddies" in the simulated volume. 
When the two scales are comparable there is an excess of correlation
within $\Lambda$, due to purely bulk motions, which could bias
high the estimated turbulent velocity.
Indeed, when we impose
$k_{\rm in}=5$ or $k_{\rm in}=1$ (tests g)-h) ) as an outer
scale to add turbulent motions in our tests, 
differences are found between
the distribution of velocity reconstructed by our method and the
correct one.  

In Fig.\ref{fig:test3} we compare the power spectra of the input velocity field and the result of our algorithm for the 
few representative cases of tests b), d), e), f), g) and h). 
In intermittent as well as uniform turbulence cases (b) and f)) the
input and measured spectra of turbulence are very similar for all scales, and they perfectly match the spectral
shape of the input turbulence at the smallest scale. 
For the case without input
turbulent field (test e)), as discussed above there is some residual pattern of misidentified turbulence. However,
these patterns contain very low kinetic energy ($\sim 10^{-3}-10^{-4}$ of the total energy), and they can be regarded as 
the unavoidable level of "noise" in our method.
As mentioned above, for turbulent modes with a scale
similar to the large-scale correlation of the profile of laminar motions (tests g-h)) our method
faces a limitation, and at turbulent power spectra at the smallest
$k$ are not fully captured.

We conclude that the method is accurate enough to separate laminar and turbulent
pattern of motions in configurations similar to the simulated ICM. 
In realistic situations, the major 
caveat to the use of our procedure is that it is difficult
to fully detect the largest-scale turbulent modes if they have a size similar to the largest scale in the computational domain. In this 
case an accurate measure of the outer injection scales of turbulent modes is only approximate, and the power spectra measured are
in general an underestimate at low $k$. This problem arises
only when the physical injection scale and the physical scale
of the ordered field are of the same order of magnitude.
We will show that this unfortunate condition
does not occur in the interesting cases of cluster mergers, cool-core sloshing and AGN-jets, which will be explored in the  remainder of the
paper.

\begin{figure*}
\includegraphics[width=0.98\textwidth]{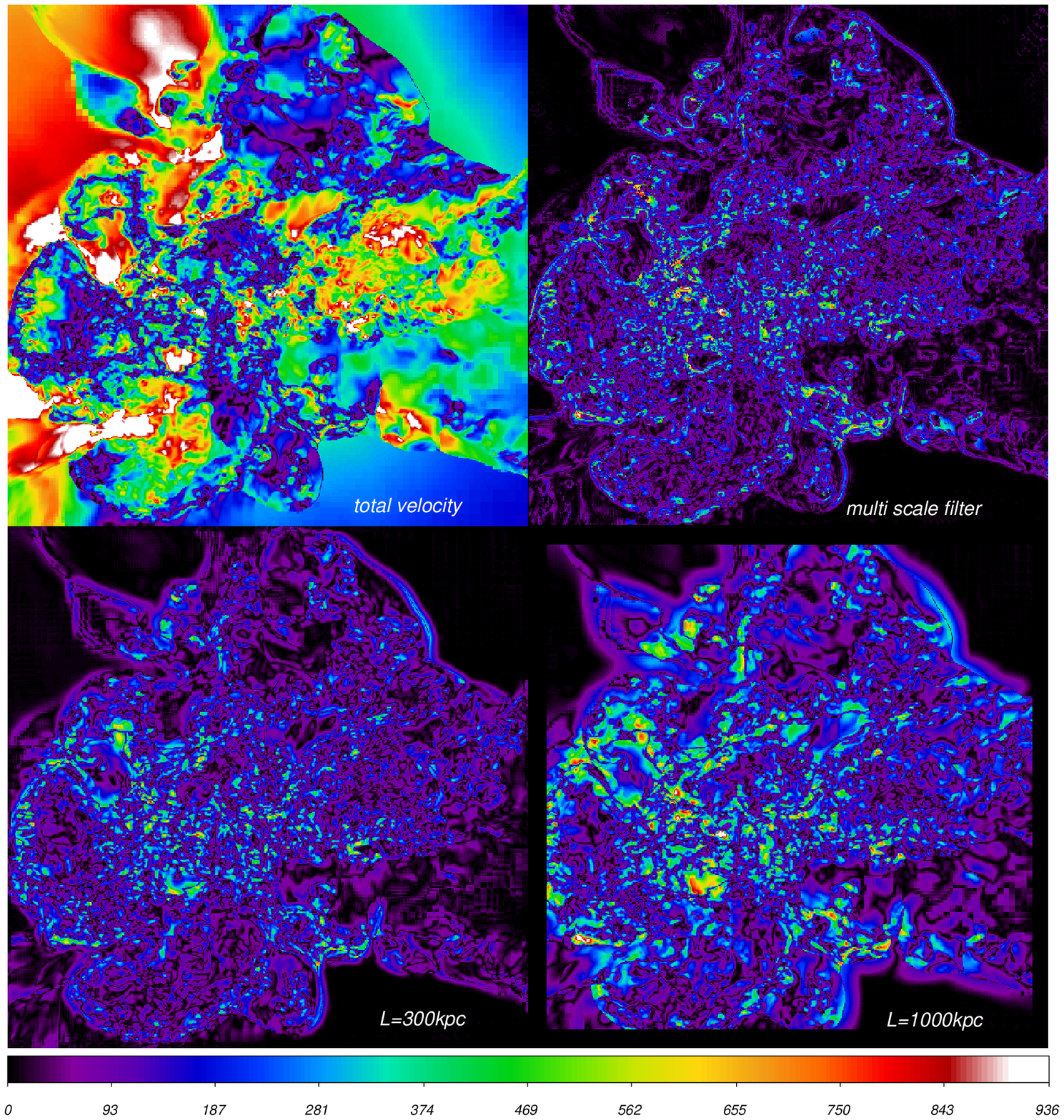}
\caption{Two-dimensional maps of total gas velocity fields through a cosmological AMR simulation at $z=0.6$. Top left: total gas velocity (in [$\rm km ~s^{-1}$]); top right: turbulent velocity field captured by our new multi-scale filter; bottom left: turbulent velocity field after the removal of $L\geq 300$ kpc scales; bottom right: turbulent velocity field after the removal of $L \geq 1000$ kpc  scales. The side of each panel is 8 Mpc$\rm ~h^{-1}$.} 
\label{fig:vel1}
\end{figure*}

\begin{figure}
\includegraphics[width=0.48\textwidth]{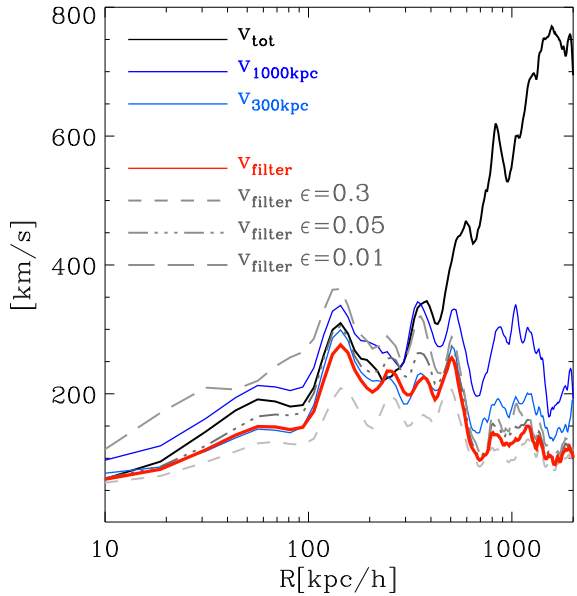}
\caption{Mass-weighted profiles of velocity from the centre of
the cluster in Fig.\ref{fig:vel1}. The different lines show the
total velocity (solid black), velocity field below the fixed
scale of 1000 kpc (blue) and 300 kpc (light blue), turbulent
velocity field reconstructed by our algorithm with fiducial parameters ($\epsilon=0.1$, in red). We additionally show in grey the results of our algorithm for different choices of $\epsilon$ to stop the iterations in Eq.\ref{eq:iter}: $\epsilon=0.3$ (dashed),  $\epsilon=0.05$ (dot-dashed) and  $\epsilon=0.01$ (long-dashed).}
\label{fig:prof1}
\end{figure}

\begin{figure}
\includegraphics[width=0.48\textwidth]{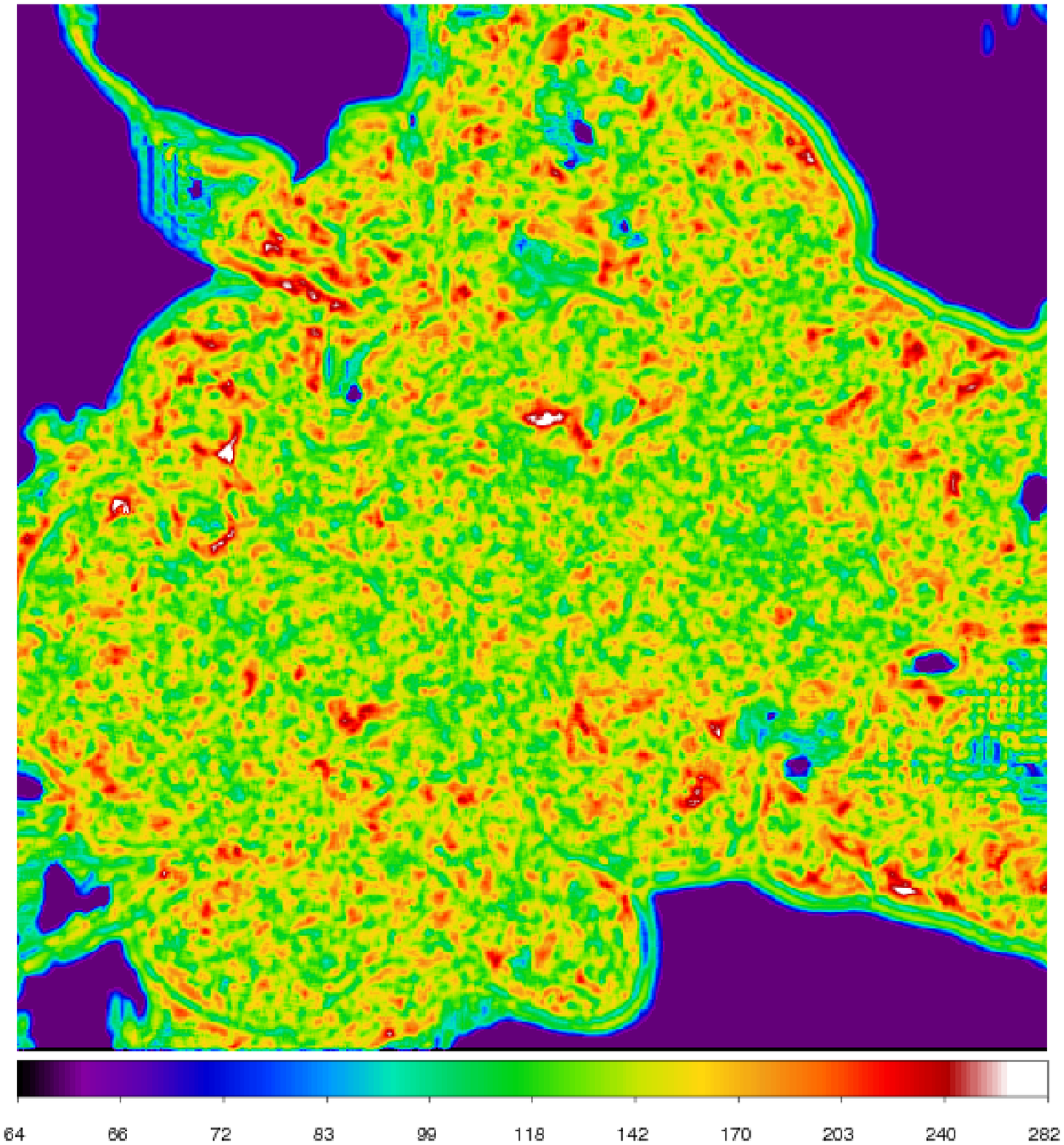}
\caption{Two-dimensional slice showing the distribution of coherence scales (in kpc) for a velocity field of the same region as in Fig. \ref{fig:vel1}.
}
\label{fig:scales}
\end{figure}


\section{Applications}
\label{sec:application}

We applied the multi-scale filter method to three important dynamical processes in galaxy clusters: 
turbulent motions in cosmological simulation of galaxy
clusters, excited by mergers and accretion
(Sect.\ref{subsec:enzo1}); turbulence in sloshing cool
cores (Sect.\ref{subsec:cool_core}); turbulence injected
by AGN outflows (Sect.\ref{subsec:agn}). In each case, we derived
the turbulent power spectra, the anisotropy of the turbulent velocity
and the resulting turbulent diffusion.

\begin{figure}
\includegraphics[width=0.44\textwidth] {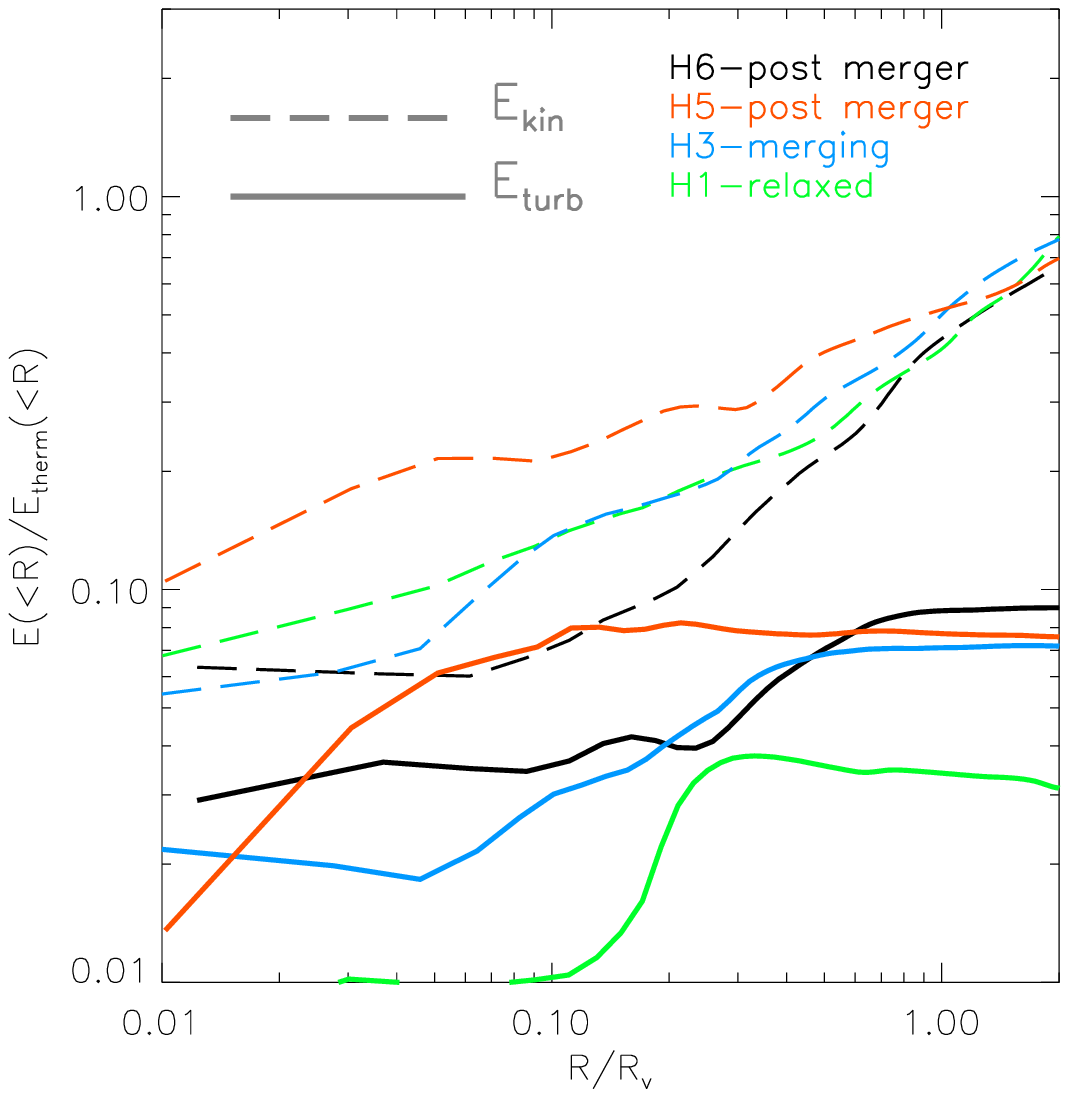}
\caption{Average radial profile for kinetic to thermal energy inside radial shells for four simulated clusters, for the total velocity field (dotted lines)
and for the turbulent velocity field reconstructed with our method (solid lines).}
\label{fig:prof2}
\end{figure}

\begin{figure}
\includegraphics[width=0.44\textwidth]{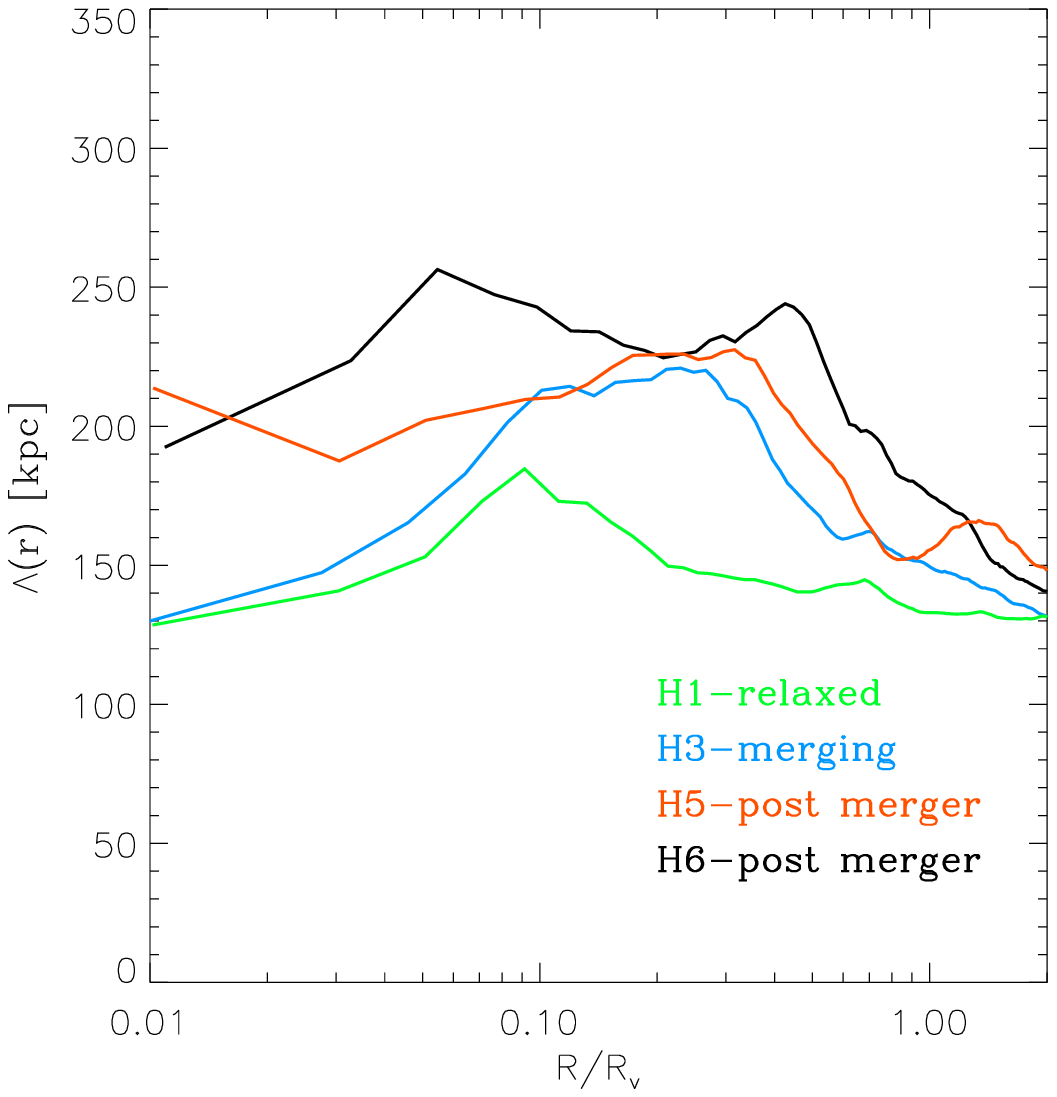}
\caption{Average radial profile of the outer scale for turbulence, $\Lambda(r)$,
for four simulated galaxy clusters.}
\label{fig:prof3}
\end{figure}


\subsection{Turbulence from mergers in galaxy clusters}
\label{subsec:enzo1}

The presence of turbulent motions on scales $\gg$ kpc in the ICM
is inferred from several observations.
Measures of Faraday rotation suggest the presence of 
chaotic super--Alfv\'enic motions in the ICM, possibly excited by merger events \citep[e.g.][]{ev03,mu04,gu08,bo10,vacca10}.
Moreover, pseudo--pressure maps of cluster cores derived from X--ray observations and the lack of  
resonant scattering effects in the X-ray spectra provide
hints of turbulence in the ICM \citep[][]{schu04,chu04,sa_fab11}.
Important constraints on the fraction of turbulent and thermal
energy in the cores of clusters are also based on the broadening
of the lines in the emitted X--ray spectra of cool-core clusters
\citep{sa10}.
In the next few years the satellite {\it Astro-H} with its high spectral resolution
will provide an important tool to observationally constrain the energy
ratio of turbulence in the ICM of real galaxy clusters \citep[e.g.][]{zhur11}.

\bigskip

Present-day cosmological numerical simulations 
routinely find that a significant
amount of pressure support (i.e. $\sim 10-30$ percent of the total 
pressure inside $0.5 R_{\rm vir}$) in the ICM is caused by 
chaotic motions, continuously excited by major and minor mergers \citep[e.g.][for recent reviews]
{iap2011sait,va11nice,jones11}.

\bigskip

Here we study turbulence in the ICM of relaxed, merging and post-merger galaxy clusters at high resolution, with the set
of simulations presented in \citet{va10kp} and \citet{va11turbo}.
These runs were produced with the  cosmological adaptive mesh refinement code {\small ENZO 1.5} \citep[e.g.][]{no07,co11}.
{\small ENZO} is currently developed by the Laboratory for Computational
 Astrophysics at the University of California in San Diego 
(http://lca.ucsd.edu).

In Fig.\ref{fig:vel1} we show the massive galaxy cluster, E1, during a major merger event (z $\approx$ 0.6).
We show  the total velocity field in the centre of
mass frame, the turbulent field after applying our multi-scale filter and the turbulent field
below the fixed filtering length of 300 kpc or 1000 kpc.

Inside the cluster atmosphere, chaotic motions are well developed and
 significantly volume filling, and similar patterns of turbulence
are detected regardless of the adopted scale for the filtering
(as long as the filtering scales is a few $\sim 100$ kpc).
This follows from the fact that, usually, the velocity field of 
the ICM in clusters is  tangled  for scales $\leq 1$ Mpc. However, the agreement
between methods using a fixed filtering scale and our method becomes less satisfactory approaching
the outer cluster regions, because towards $R_{\rm vir}$
large-scale infall motions become more frequent, and large patches
of laminar infalling gas are found in correlations
with large-scale filaments. 
In these cases, equally strong smooth and chaotic flows can be found
at roughly the same distance from the cluster centre,
and distinguishing among them is quite difficult if 
using a fixed scale.

We compare in Figure \ref{fig:prof1} the mass-weighted radial profiles
of total velocity and several estimates of turbulence in the same volume: by assuming fixed filtering scales (300 kpc and 1000 kpc) and
with our multi-scale filtering algorithm. In the same figure, we additionally show the results of slightly different choices of $\epsilon$ to stop the iterations in Eq.\ref{eq:iter} ($\epsilon=0.3$, $=0.05$ and $=0.01$).
Both fixed filtering scales yield in general a slightly higher turbulent velocity at all radii compared to our method; the differences
with respect to the filter$=300$ kpc are on average very small. On the other hand, the turbulent velocity estimated with our method is lower by a factor $\sim 2-5$ , with respect to the total velocity field of this cluster, for $R>500\rm ~kpc~ h^{-1}$. 
The choice of different values of $\epsilon$ to stop the iterations
of our algorithm does not have dramatic consequences in the reconstructed velocity field. However, differently from the  
our tests in Sec.\ref{subsec:test}, in this more realistic situation the choice of very low values of $\epsilon$ may also cause problems,
because to pin down the fractional change of the turbulent
velocity field around the cell, very a large volume is scanned in the
iterations, which can be as large as the cluster itself. Our fiducial
choice of $\epsilon=0.1$ ensures a reasonable compromise between 
accuracy in the iterations, and the need of avoiding contamination
from well-separated regions (and shocks) in simulated galaxy clusters.

For completeness, we also show in Fig.\ref{fig:scales} a
map of the scale $\Lambda$ 
of the velocity field inside the same scale as for the same region of Fig.\ref{fig:vel1}. 
The value of $\Lambda$ shown is the average between the three velocity components. 
Across most of the cluster volume, $\Lambda$ fluctuates in the range $\sim 100-300$ kpc. This suggests that, on average, the use
of a simple fixed filtering scale in this range (as performed in the past by  \citealt[][]{do05,va06,va09turbo}) is still a good approximation
to study turbulent motions in the ICM, and that the energy profiles
obtained with this technique in the past are fairly consistent
with these new and more elaborated ones.

\begin{figure}
\includegraphics[width=0.48\textwidth]{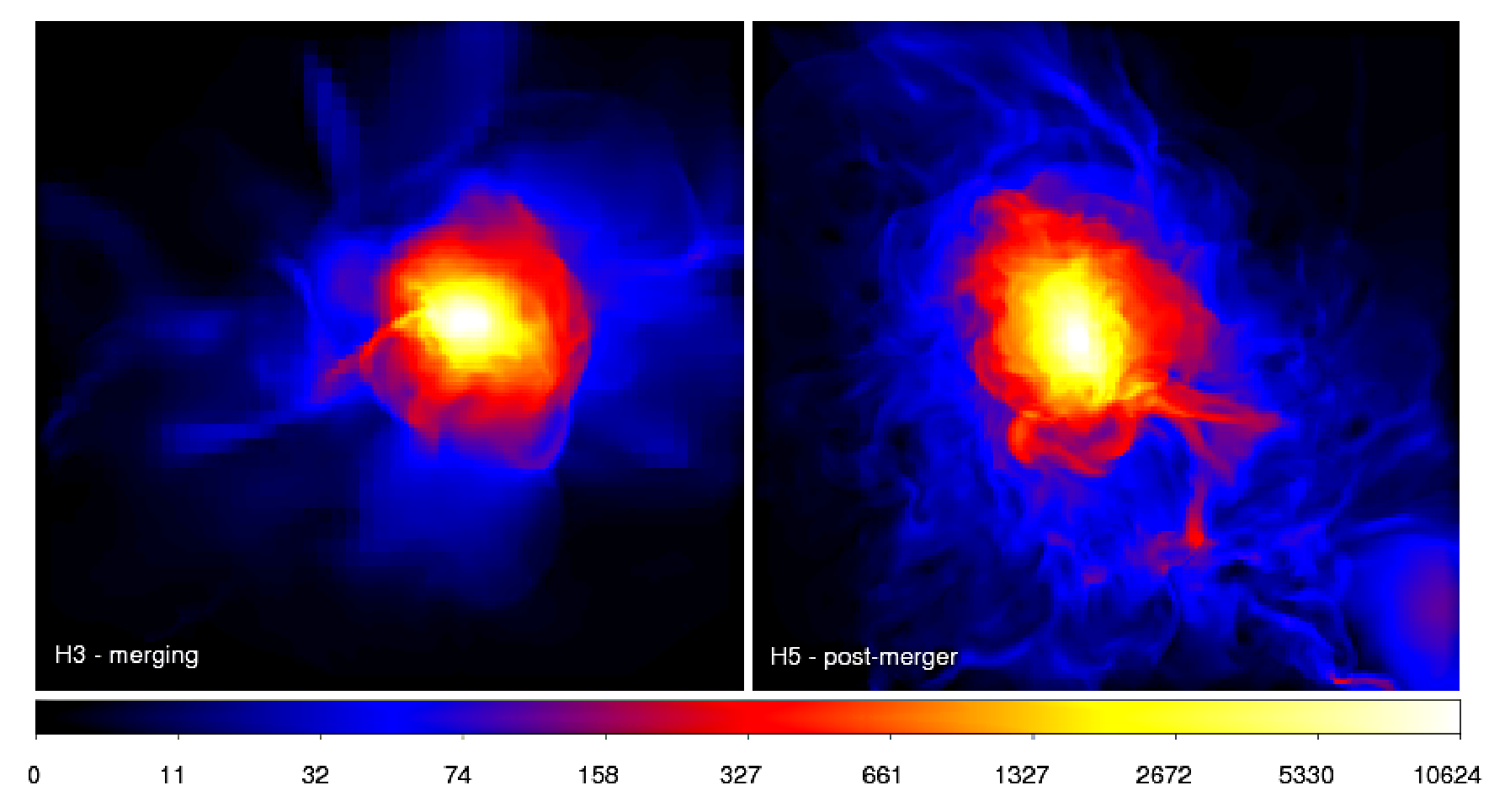}
\includegraphics[width=0.48\textwidth]{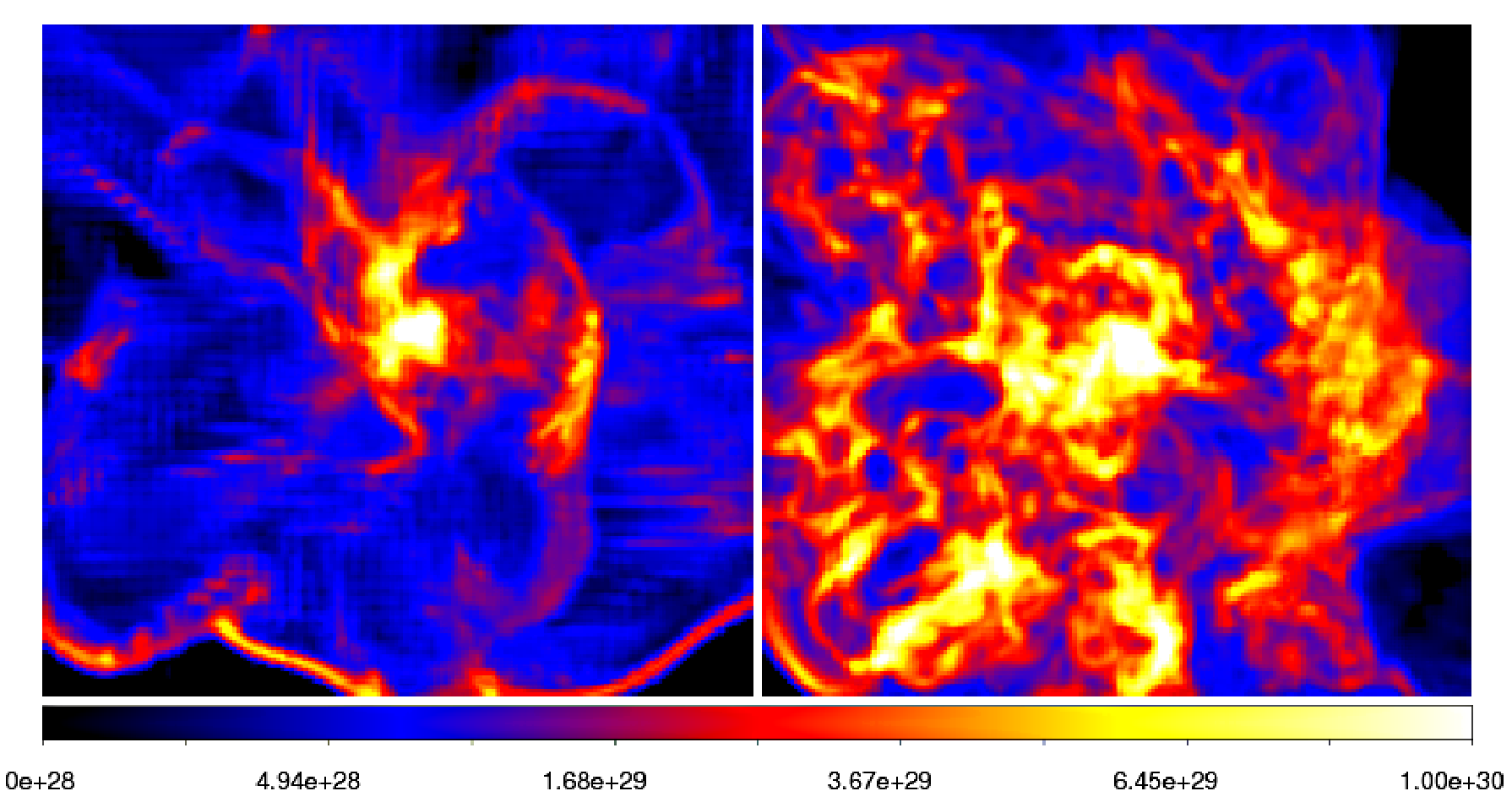}
\caption{Top panel: map of gas density for a slice of $100 \rm~kpc~h^{-1}$ through the
centre of the major merger cluster H5 (right column) and of the merging cluster H3 (left column). The top row shows the projected average gas density (in [$\rho/\rho_{\rm cr,b}$], where $\rho_{\rm cr,b}$ is the critical baryon density), the bottom row shows the projected map of the turbulent
diffusion for the same regions (in units of [$\rm cm^{2}~s^{-1}$]).}
\label{fig:diffusion}
\end{figure}

\begin{figure}
\includegraphics[width=0.44\textwidth]{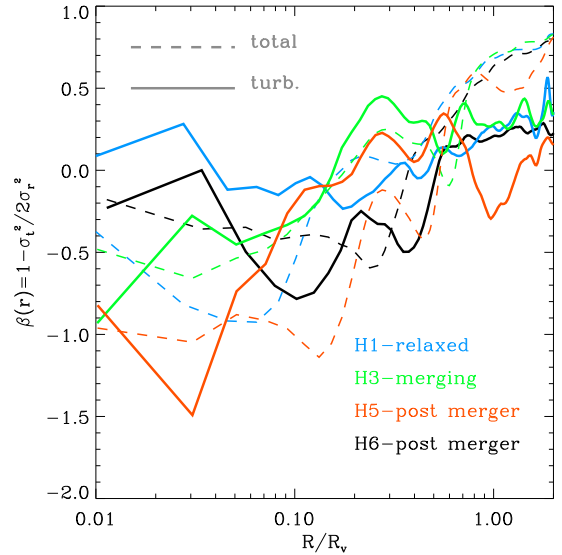}
\caption{Average radial profile of the anisotropy parameter (for the total
velocity field, in dashed, or for the turbulent velocity field, in solid)  for four simulated galaxy clusters.}
\label{fig:prof5}
\end{figure}

\subsubsection{Turbulent energy budget in clusters}

We applied our filtering procedure to four additional galaxy clusters of total final virial mass $M_{\rm tot} \sim 3 \cdot 10^{14} M_{\odot}~ \rm h^{-1}$,  which we
already studied in previous works \citep[][]{va10tracers}.
These systems have different dynamical states: we have two "post-merger" objects (H5 and H6,  with a merger with a mass ratio higher than $1/3$ for $z \leq 0.5$), one merging
cluster (H3) and one relaxed system (H1, without evidence of past or ongoing major merger for $z<0.5$).
Figure \ref{fig:prof2} shows the average radial profiles of
turbulent and total kinetic energy within shells for these clusters, normalised to the thermal energy within the same volume,
similar to Fig.\ref{fig:prof1}. 
The turbulent energy of cells is computed as $E_{\rm turb}= \Delta x^{3} \rho  \sigma_{\rm v}^{2}/2$, where $\sigma_{\rm v}$ is the modulus of the 3D velocity field below the assumed
spatial scale, computed in Eq.\ref{eq:turbo_vel}, and $\Delta x$ is the resolution of the cell.
Using the total velocity field would overestimate the turbulent budget by a factor $\sim 3-10$ at all radii for all objects. The ratio between turbulent and thermal energy flattens with radius in the range $\sim 0.5-2 R_{\rm vir}$,
with $E_{\rm turb}/E_{\rm therm} \sim 0.1$,  while the kinetic energy increases continuously with radius and approaches
the thermal energy budget at $R_{\rm v}$. The trend of $E_{\rm turb}/E_{\rm therm }$ with the dynamical state of host clusters is quite
regular inside $\sim R_{\rm vir}$ (post-merger systems present a higher content of turbulent energy compared to relaxed systems, while merging systems stay in between), whereas for $\leq 0.5 R_{\rm vir}$
the trend becomes sensitive to the timing of the merger event, and to shock heating episodes
(which affect the thermal energy of the ICM).

The radial turbulent energy weighted profiles of 
the maximum coherence scale, $\Lambda(r)$, for these same clusters are shown in Fig.\ref{fig:prof3}. 
The trend of this scale with cluster dynamical states suggest that on average the most perturbed systems host the largest turbulent
patterns, with $\Lambda \sim 250$ kpc, while the average values are almost half of that are found for the relaxed system H1.
The differences tend to be smaller at $R_{\rm vir}$, where on average all systems present correlation scales of 
$\Lambda(r) \sim 100-150$ kpc.

 These results cast
doubts on the usual assumption of an injection scale of turbulence at shocks of the order of the
curvature radius of accretion shocks, $\sim 0.5-1 ~R_{\rm v}$ \citep[e.g.][]{ca11}. 
For our clusters this would indeed imply values of  $\sim 1-2 \rm~ Mpc~ h^{-1}$
for the outer scale of turbulent motions, $\Lambda$,  one order of magnitude larger than what we measure here.
Also, the energy budget at $\sim R_{\rm vir}$ would be overestimated by 
a factor $\sim 5-10$ by assuming such a high value of $\Lambda$. The reason for this significant difference is likely that,
in the outskirts of simulated galaxy clusters, turbulence is mostly injected at the scale $\sim 100-300 \rm~kpc$, typical
of the density/pressure inhomogeneities of the ICM, downstream of accretion shocks.

\begin{figure}
\includegraphics[width=0.49\textwidth]{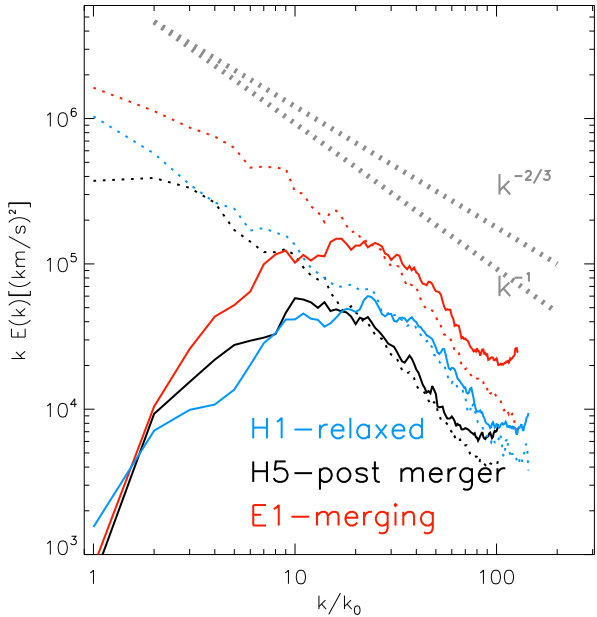}
\caption{Energy spectra of the 3D velocity field for three galaxy clusters (E1, H1 and H5). The dotted lines show the spectra of
the total velocity field, the solid lines show the spectra of
the turbulent velocity field. The gray lines
shows the slope of $\alpha=1$ and $\alpha=2/3$ to guide the eye.}
\label{fig:pk}
\end{figure}

\subsubsection{Turbulent diffusion in clusters}

Our algorithm offers a straightforward estimate of turbulent diffusion in the simulated ICM, that is of the order of:

\begin{equation}
D_{\rm turb} \approx 0.11 \cdot \Lambda_{\rm I} \cdot \sigma_{\rm v},
\label{eq:difussion}
\end{equation}

\noindent where $\Lambda_{\rm I} \approx 9 \Lambda/20$ \citep[see for instance][]{dc05}). Together
with other mechanisms in the ICM, such as stellar feedback
or galactic winds, turbulent
diffusion may have an important role in transporting
metals from active galaxies to the ICM. For instance, \citet[][]{re06}
analysed the diffusion coefficient needed to model the 
metallicity observed around several central galaxies in nearby
clusters, and found most likely values of the turbulent diffusion 
in the range $D_{\rm turb} \sim 10^{28}-10^{29}  {\rm cm^{2}~s^{-1}}$. 
 
We show in Fig.\ref{fig:diffusion}
the projected map of volume-weighted turbulent diffusion
for the post-merger system H5 and for the merging cluster H3 along with their projected density for a slice of depth 200 kpc.
In the major merger cluster H5 the ICM is characterised by volume-filling turbulent motions, which attain a maximum of
$D_{\rm turb} \sim 0.5-1 \cdot 10^{30}  {\rm cm^{2}~s^{-1}}$  in localised patches of a few $\sim 100$ kpc in size, and $\sim 10^{29} {\rm cm^{2}~s^{-1}}$ elsewhere. In the pre-merger system H3 most of the cluster volume is characterised by lower values of turbulent diffusion, $D_{\rm turb}<5 \cdot 10^{28} {\rm cm^{2}~s^{-1}}$, but a few localised patches attaining higher diffusion are found related to minor mergers along the direction of the large-scale
infall of matter (E-W direction in the image).  Values in the same range are measured in the virial volume of the other two clusters.

These results agree with the estimates of 
turbulent diffusion of previous studies of the
simulated ICM using tracers \citep[][]{va10tracers}
and imply that the efficiency of particle transport in these
simulated clusters is much higher than that caused by thermal diffusion \citep[e.g.][]{sht10}.

 In the following sections we compare the distribution of turbulent diffusion in these runs
with that of cool-core sloshing in the Virgo cluster (\ref{subsec:cool_core}) and with that of AGN outflows (\ref{subsec:agn}).

\begin{figure*}
\begin{center}
\includegraphics[width=0.9\textwidth]{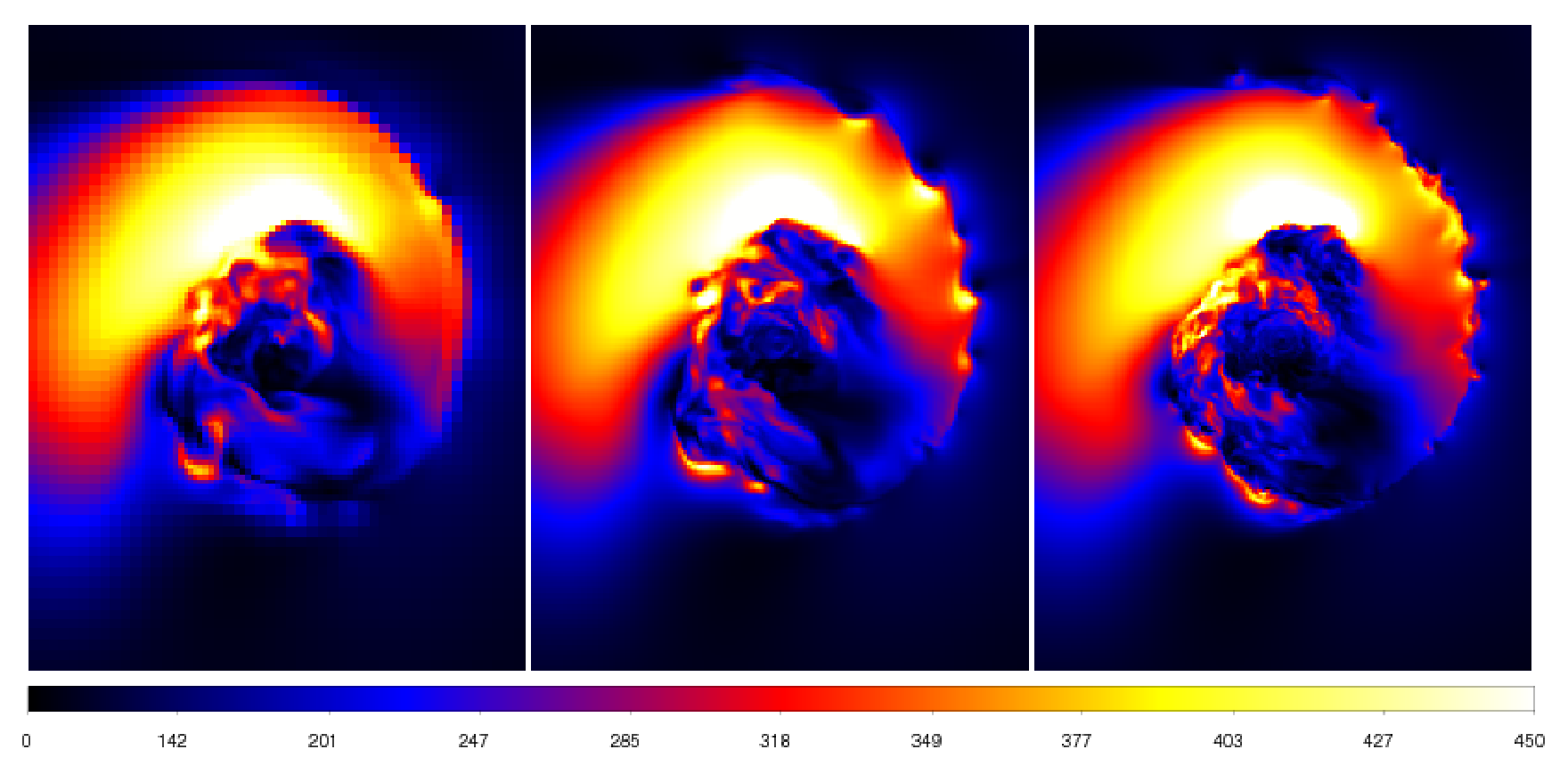}
\includegraphics[width=0.9\textwidth]{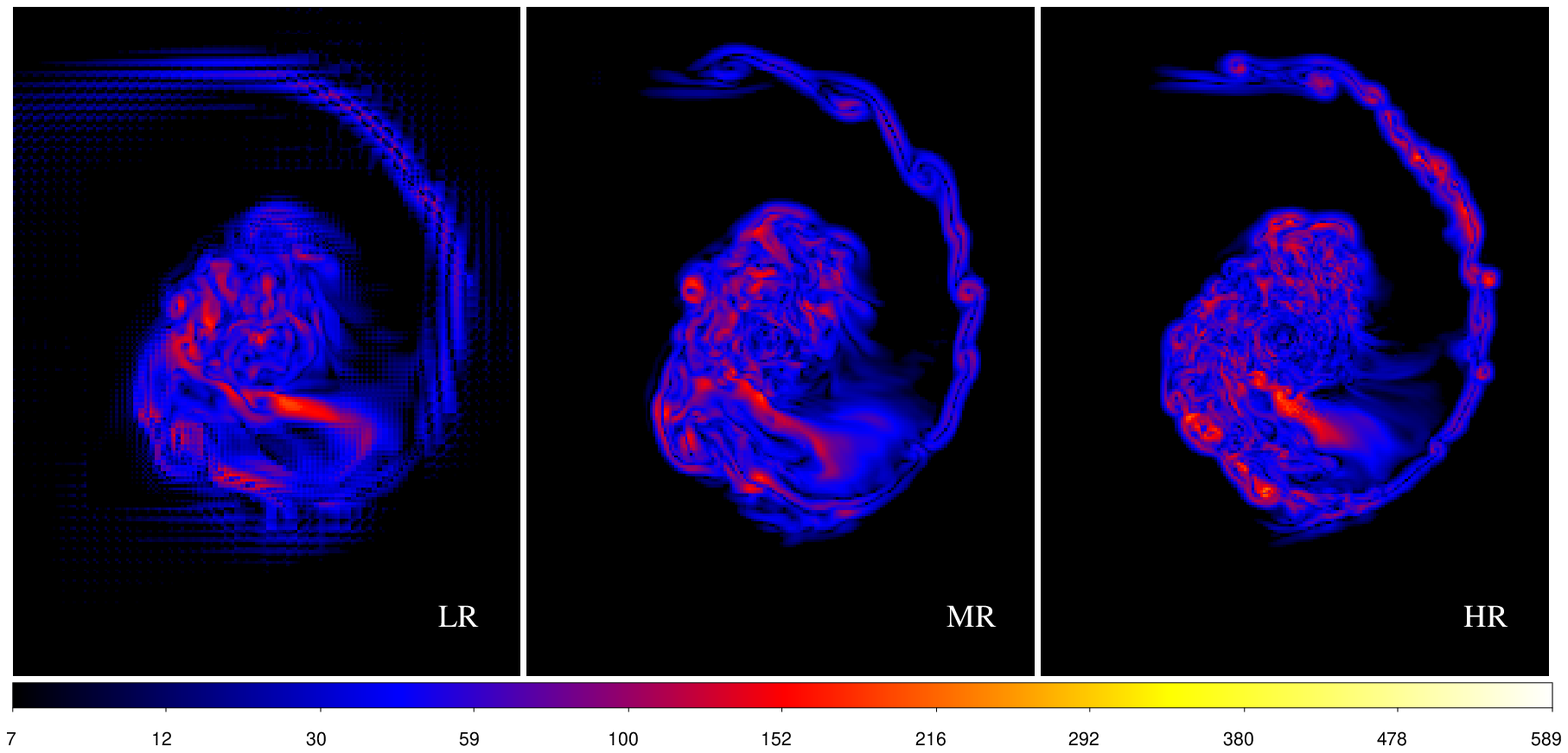}

\caption{Gas sloshing in the VIRGO cluster, triggered by a minor merger \citep[][]{roe11}.
The images are slices through the centre  of the simulation box, showing the absolute value of the total (top panels) and 
turbulent (lower panels) velocity for the three different resolutions: 2 kpc (LR), 1 kpc (MR) and 0.5 kpc (HR).  
The colour coding is in [$\rm km~ s^{-1}$], each image has sides $\sim 250 \times 300$ kpc.}
\label{fig:virgo1}
\end{center}
\end{figure*}

\subsubsection{Anisotropy of turbulence in clusters}

Our method also offers a way to monitor the anisotropy of
the turbulent velocity field within the cluster volume, as a
function of radius,

\begin{equation}
\beta(r)=1-\frac{\sigma_{\rm v,tan}^2}{2\sigma_{\rm v,rad}^2},
\end{equation}

\noindent where the turbulent field is decomposed into its tangential
and radial component from the cluster centre. Knowing the
anisotropy of the velocity field of galaxy clusters simulated
in "simple physics" cosmological simulations is important to
pinpoint the additional effects of MHD instabilities, which
can potentially lead to a radial alignment of $\vec{B}$ and
of the velocity field in the cluster outskirts \citep[e.g.][]{quat08,ruszkowski11,parr11}. 
Without a reliable method of detecting turbulent motions,
laminar contributions of large-scale bulk flows can significantly
bias the estimate of radial turbulent motions in the ICM.

Figure \ref{fig:prof5} shows the average profiles of $\beta(r)$ for the four clusters; as a comparison we also overplot
the corresponding profiles of the anisotropy parameter that
we would obtain
from the total unfiltered velocity field. Close to $R_{\rm vir}$, we note
the striking feature that while the total
velocity field is preferentially radial ($\beta \sim 0.5-1$) , the turbulent velocity field is close to isotropic
($\beta \sim 0$). Inside the virial radius,
both velocity fields become preferentially tangential.
Also in this case, the turbulent velocity field of all clusters
is in general closer to isotropic than the
total velocity field. These trends follow from the fact that while the bulk motions of satellites are characterised by strong
radial motions towards the cluster centre, ram pressure stripping and the  hydro-dynamical interaction with the ICM 
inject chaotic motions in a more isotropic way inside $\sim R_{\rm vir}$. 


\subsubsection{Power spectra of turbulence in clusters}

Finally, we compute the power spectra of the turbulent velocity field and  compare them with the
power spectrum of the unfiltered velocity field studied in our previous works \citep[][]{va10tracers,va11turbo}.
In Fig.\ref{fig:pk} we show
the spectra for three clusters (E1 at z=0.6, H1 and H5 at z=0)
 with a zero-padding technique and employing an apodisation
 function to avoid spurious effects at the edges of the domain \citep[][]{va10tracers,valda11}. The wavenumbers are normalised to the virial radius of each cluster, $k_{0}=2 \pi/R_{\rm vir}$. To determine the injection scale, for each cluster we plot the spectral energy per mode,
 $k \cdot E(k)$ (a Kolmogorov spectrum would have a $\propto k^{-2/3}$
 scaling here, see the dotted gray lines in the Figure). 
Unfiltered spectra (dotted lines) present 
power law spectra for $k>1-2 ~k_{0}$, with slopes in the range of $\alpha \sim 2/3 - 1$. For the post-merger cluster
H3, there is some flattening on scales
of the virial radius. These results qualitatively
agree with power spectra reported in the literature, based on different numerical methods \citep[][]{do05,ryu08,xu09}.

The turbulent velocity spectra of all clusters show a peak at $k/k_{0} \sim 10-30$, corresponding to spatial scales $\sim 0.1-0.3 R_{\rm vir}$ ($\sim 150-500 \rm ~ kpc~h^{-1}$ for the range of masses we consider here).  We cannot detect a single sharp scale responsible of the injection
of turbulent energy in the cluster volume, but a quite broad range of scales,  
consistent with the patchy
distribution of scales shown in Figs.\ref{fig:scales} and \ref{fig:prof3}. For the post-merger system H5, the turbulent energy peaks at a larger spatial scale compared to the relaxed cluster of equal mass, H1, implying the presence 
of strong turbulent motions caused by the most recent
merger event. At smaller spatial scales, $k/k_{0}> 50$ ($\leq 300 \rm kpc~ ^{-1}$), all
clusters show a power-law spectrum with a slope slightly steeper than the Kolmogorov one.

Our findings suggest that, despite the clear
power-law behaviour of the spectral energy distribution
of the ICM velocity field (which runs for almost 2 orders
of magnitude in spatial scales), turbulent motions
indeed dominate the cascade of energy only for $k/k_{\rm 0}>30$ (corresponding to $\sim 0.3 R_{\rm vir}$, or $\sim 0.5-1\rm~Mpc~ h^{-1}$ for these masses). 
The spectral behaviour of the 3--D velocity field of the ICM for scales $\geq 0.3-0.5 R_{\rm vir}$ is on the other hand dominated by 
the pattern of velocities driven by large-scale infall, whose
kinetic energy is mostly characterised by a laminar pattern.

\subsection{Cool core sloshing}
\label{subsec:cool_core}

Our next example concerns ICM sloshing and the resulting formation of cold fronts (CFs). These are discontinuities in X-ray brightness and temperature, where the brighter side is also the cooler one. They come in two varieties (see also review by \citealt[][]{mv07}): {\it merger CFs} with stronger temperature contrasts across the fronts are the contact discontinuities between the intra-cluster media of two merging clusters, and {\it sloshing CFs} \citep[][]{mvm01} are named after their most likely origin. For the latter, the idea is that a gas-free subcluster moved through a galaxy cluster and its ICM. During the pericentre passage, the combined gravitational and hydrodynamical interaction slightly offsets the ICM in the cluster core without disrupting it. After the subcluster has passed the central region and moves away, the offset ICM falls back towards the main cluster DM peak and starts to slosh inside the main potential well \citep{am06}. Thus, sloshing CFs are contact discontinuities between gases of different entropy, originating from different cluster radii. Usually, the subcluster passes the main cluster core at some distance, it transfers angular momentum to the ICM, and the sloshing takes on a spiral-like appearance, and so do the resulting CFs, which are wrapped around the cluster core {\footnote{This type of CF is reported to be ubiquitous \citep{mv07}, and high-resolution observations are available for several clusters \citep[see][and references therein]{roe11}.}}. Measuring the amount of small-scale turbulent motions in these 
simulations is important because the excitation of turbulence around
sloshing cool cores has recently been proposed as 
a mechanism to power radio mini-halos via turbulent re-acceleration
of $\gamma \sim 10^{3}$ electrons in the
magnetised ICM \citep[e.g.][]{mg08,zu11sait}.

\citet[][]{roe11} simulated ICM sloshing specifically in the Virgo cluster, using the AMR code {\small FLASH} (version 3.2  \citealt{dubey09}). The simulations were performed in 3D in a simulation box of size of $3 \times 3.5 \times 3 {\rm Mpc}^{3}$.

Besides the superimposed sloshing and rotational large-scale motions, hydrodynamical instabilities at the CFs introduce a certain amount of turbulence, which we analyse here for the fiducial case with a subcluster of mass $2 \cdot 10^{13} M_{\odot}$, scale radius 100 kpc, and pericentre distance of 100 kpc. 
To check the dependence of the turbulent velocity field on the numerical resolution, we compared three re-simulations of the Virgo cluster with increasing
maximum resolution: $\Delta x \approx 0.5$ kpc (HR run), $\approx 1$ kpc (MR) and $\approx 2$ kpc (LR).  

Figure \ref{fig:virgo1} shows the trend with
resolution of the total velocity
field (top panels) and turbulence (bottom panels)
for thin cuts through the middle of runs LR, MR, and HR at the same time step. To
run our algorithm over the same number of cells, we sampled all outputs at the resolution of the LR run (2 kpc).

The increase
of resolution enables us to capture the formation
of Kelvin-Helmholtz (KH) instabilities along the spiral arms in detail, and to
separate them more efficiently from the large
scale rotation. 
The coherent rotation is characterised by values of $\sim 300-500 \rm ~km~ s^{-1}$ at large scale in all runs.  At low resolution (2 kpc) no KH rolls are observed along the spiral arms of the sloshing core, 
while at intermediate (1 kpc) and high resolution (0.5 kpc) the
KH instabilities can form, and develop patches of turbulent velocities of up to $\sim 100-200\rm~km~ s^{-1}$. 

\subsubsection{Power spectra of turbulence in sloshing cool cores}

\begin{figure}
\includegraphics[width=0.45\textwidth,height=0.4\textwidth]{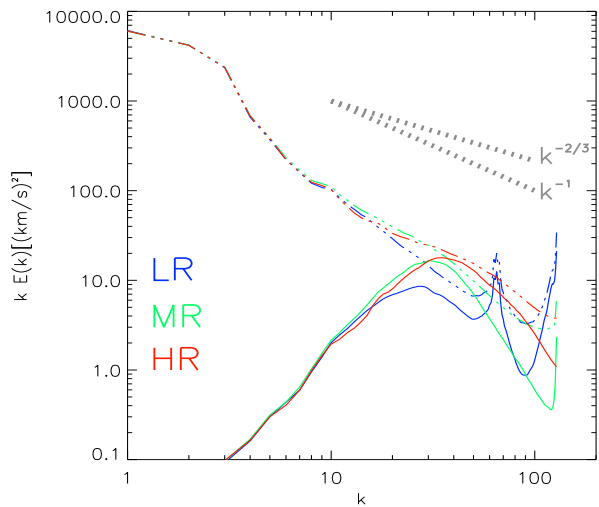}
\caption{Power spectra for the total velocity field (dot-dashed lines)
and for the turbulent velocity field (solid lines) for the 
Virgo simulation at the three resolutions. The 
spectra are computed for a cubic region with the side of 250 kpc.
The x-axis is in unit of $2 \pi/L$, where $L$ is the size of the box.
The features at $k \geq 50$ only present in run LR are an artifact of coarse
resolution.}
\label{fig:virgo3}
\end{figure}

In Fig.\ref{fig:virgo3} we present the 3D power spectrum
of the total and of the turbulent velocity field for a region
of $(500 {\rm kpc})^{3}$ in the three runs. 
A large-scale correlation in the total velocity is found at
all resolutions, with an overall slope steeper than the Kolmogorov
spectrum. This correlation is mostly due to the
correlation imposed by the large-scale pattern of rotation after
the crossing of the cluster satellite.

The increase of resolution creates a bump in the turbulent velocity spectrum
for $k>30$ (corresponding to scales $<20$ kpc); this bump exactly
corresponds to the peak of turbulent energy reconstructed
by our algorithm, suggesting that the increase of resolution causes a real development of turbulent motions in this range
of scales.

\subsubsection{Anisotropy of turbulence in sloshing cool cores}

\begin{figure}
\includegraphics[width=0.45\textwidth,height=0.4\textwidth]{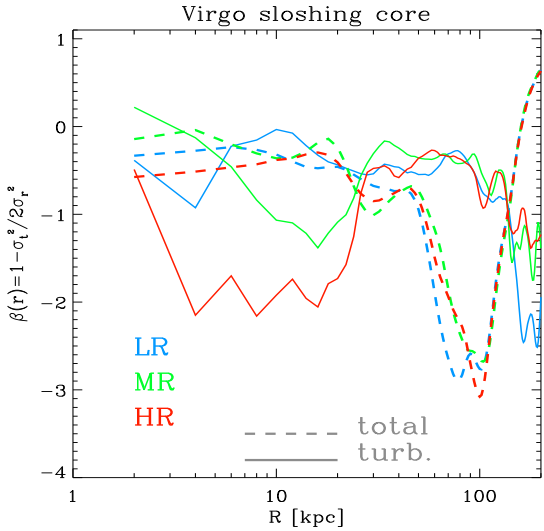}
\caption{Radial average profiles of the anisotropy parameter, $\beta(r)$, for the Virgo simulation at three resolutions. The 
dashed lines are for the total velocity field, the solid lines
are for the turbulent velocity field.}
\label{fig:ani_virgo}
\end{figure}

Furthermore, we computed the average
radial profiles of the anisotropy parameter for the total and the
turbulent velocity field in these three runs (Fig.\ref{fig:ani_virgo}). However, in contrast
to our previous results on clusters,
the tangential motions of the sloshing core strongly dominate
the total velocity field inside 100 kpc, and become more
isotropic approaching the innermost cluster core. The convergence
on $\beta(r)$ is very good at all radii $>20$ kpc.
The turbulent field is also preferentially tangential inside 100 kpc, but no clear convergence with resolution is
observed within the cluster core, likely in response to the growth of KH instabilities at small scales as resolution is increased.
In the top panels of Fig.\ref{fig:virgo_component} we show the
components of the turbulent velocity field for a slice of 1 kpc
in run HR. 
The turbulent motions are well-confined within the
innermost 100 kpc of the cluster, also when seen in projection,
as shown in the bottom left panel.

\subsubsection{Turbulent diffusion in sloshing cool cores}

\begin{figure*}
\includegraphics[width=0.94\textwidth]{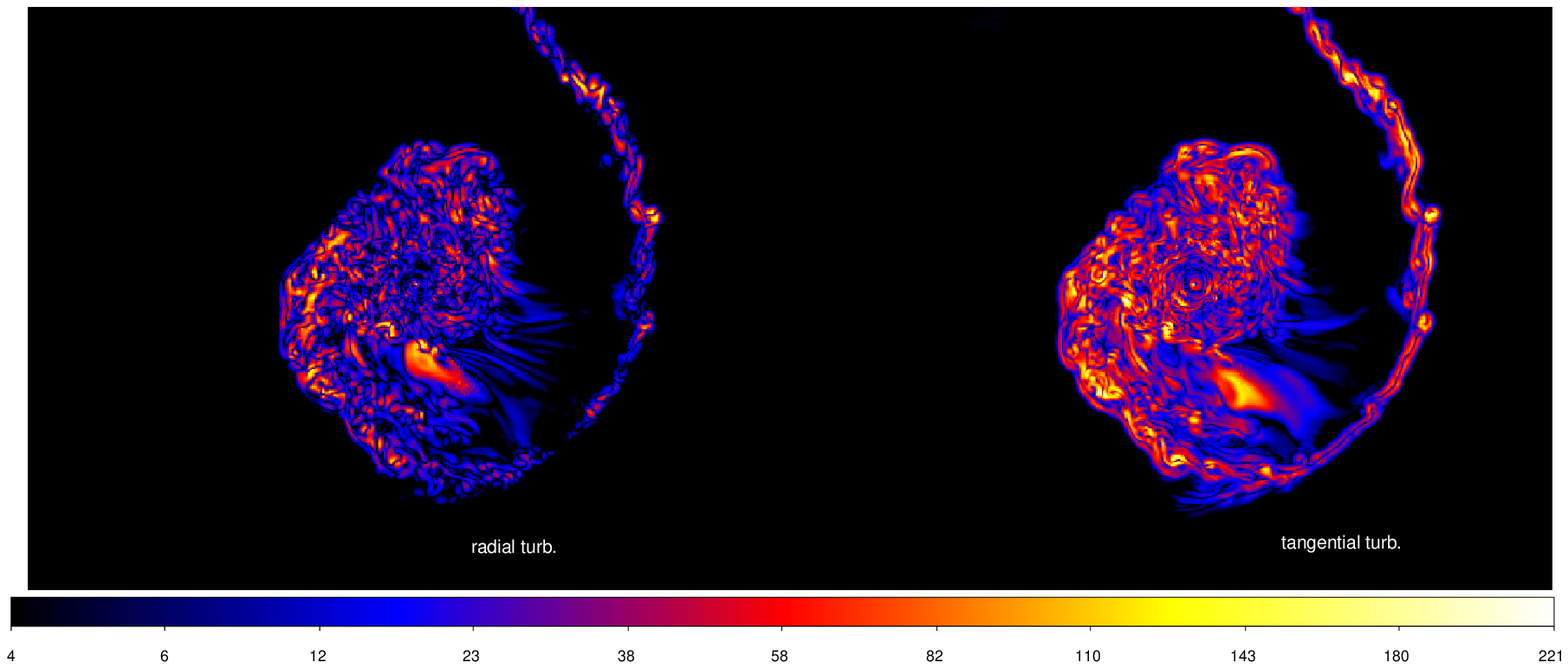}
\includegraphics[width=0.47\textwidth,height=0.47\textwidth]{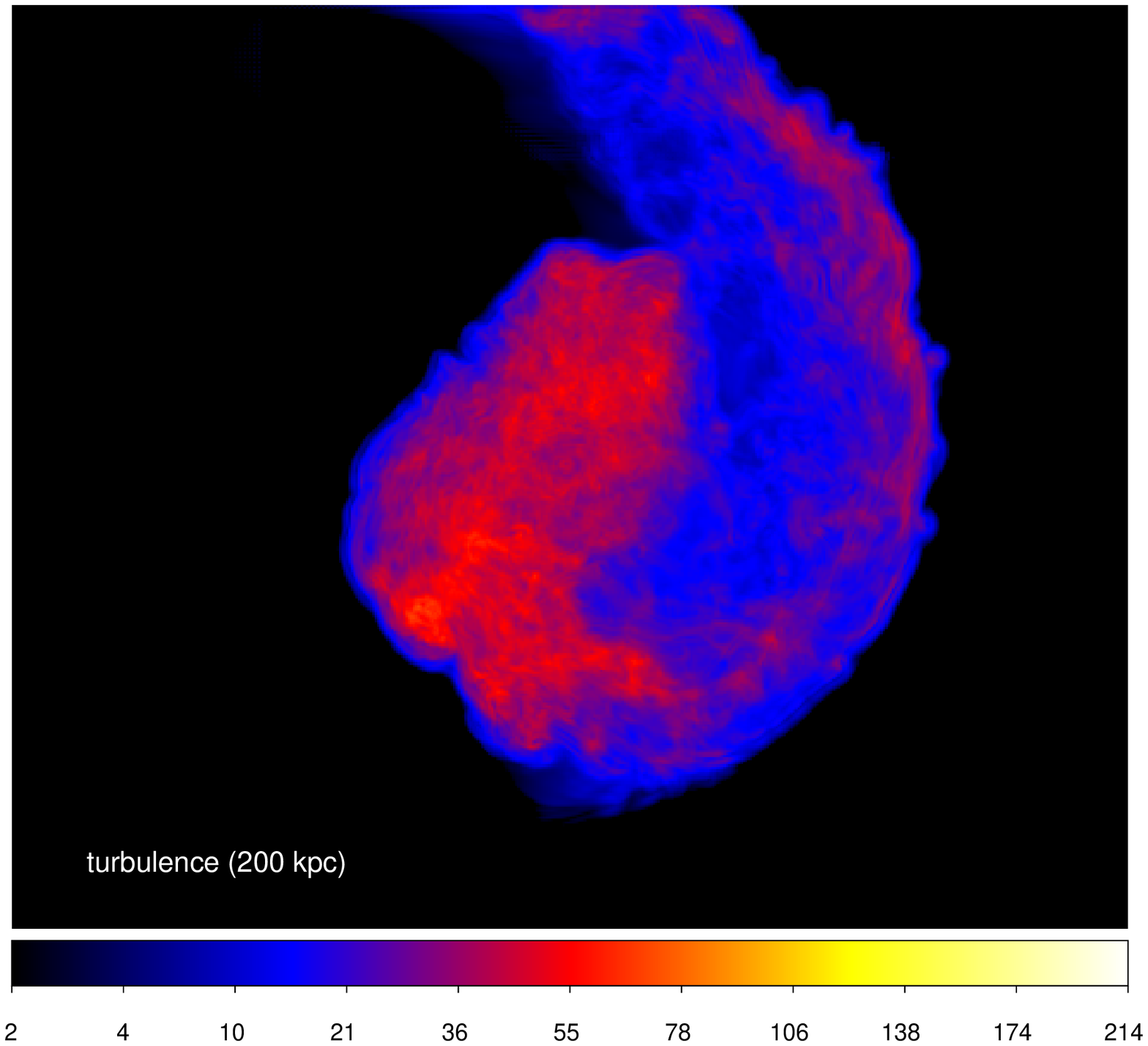}
\includegraphics[width=0.47\textwidth,height=0.47\textwidth]{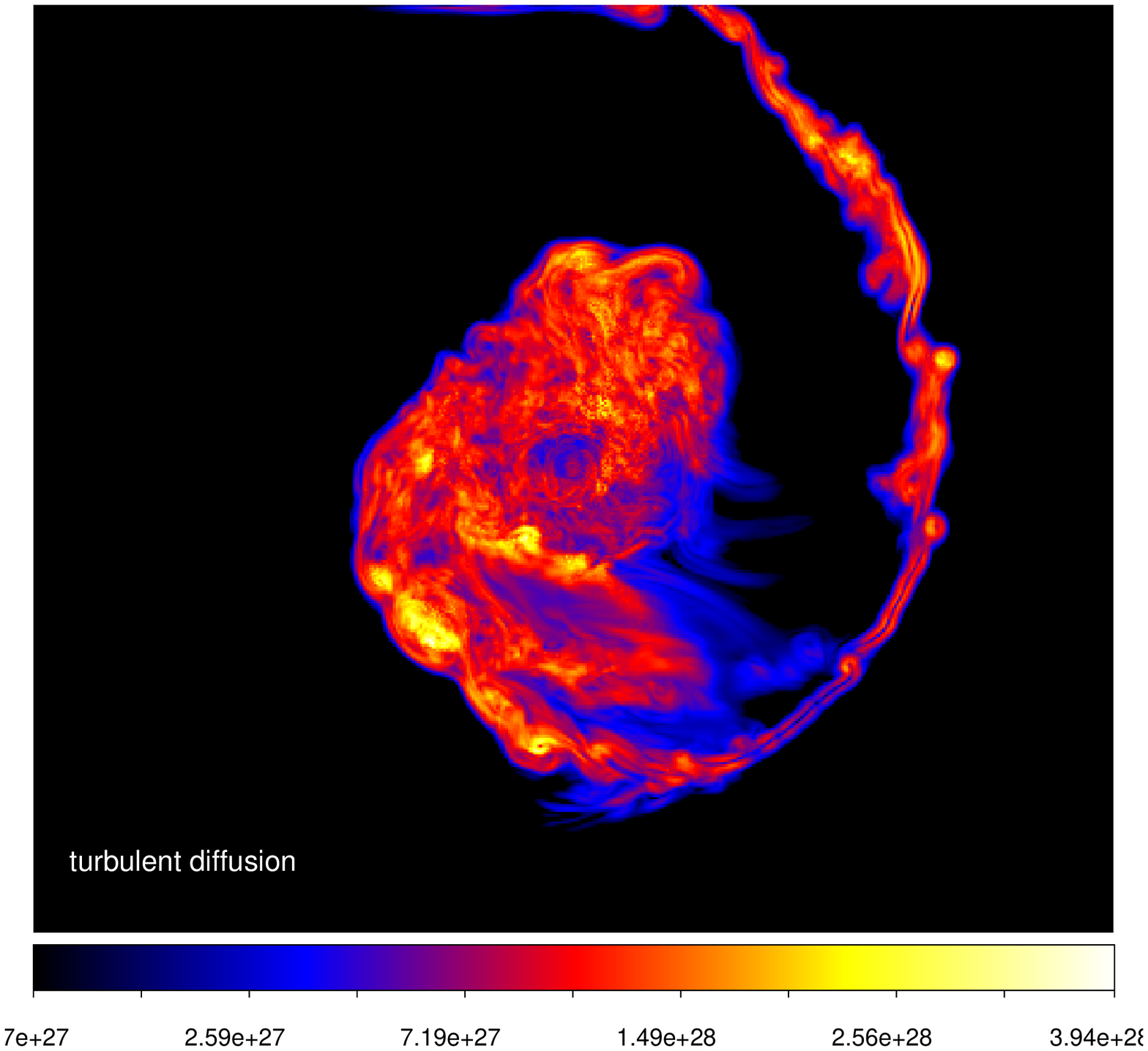}
\caption{Top panels: slices of 1 kpc through the centre of the Virgo run HR, showing the radial component of the turbulent velocity field (left, in [km/s])
and the tangential one (right,in [km/s]). Bottom panels: volume-averaged maps of turbulent velocity field (im [km/s]) for a line of sight of 200 kpc in run HR (left panel) and maps of turbulent diffusion for a slab of 20 kpc (right, in units of [$\rm cm^{2}~ s^{-1}$]). Each image has sides $400 \times 500$ kpc.}
\label{fig:virgo_component}
\end{figure*}

As for the case of clusters, we computed the turbulent
diffusion reconstructed with our method (bottom right panel
of Fig.\ref{fig:virgo_component}). Values of 
$D_{\rm turb} \sim 1-2 \cdot 10^{28} {\rm cm^{2}~ s^{-1}}$ are found in
a thin stripe along the spiral arm of the sloshing, and in the
innermost $100$ kpc around the cluster centre. It is intriguing
that despite the differences in resolution and driving 
mechanism in the simulations, this range of values is very similar to what we obtained for the case of galaxy clusters simulated with {\it ENZO} (Sec.\ref{subsec:enzo1}).

\begin{figure}
\includegraphics[width=0.48\textwidth]{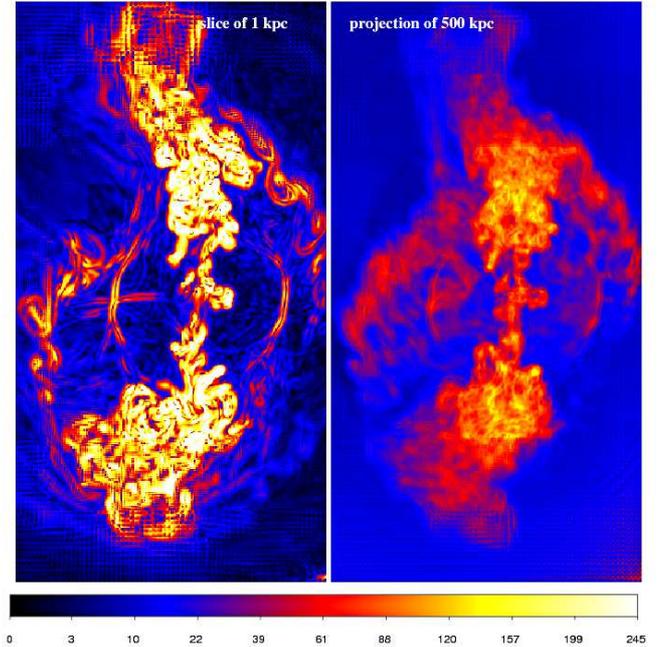}
\caption{Maps of turbulent velocity module (in units of [$\rm km~ s^{-1}$]) for the a slice of 1 kpc 
centred on the Hydra run (left) and for the volume-weighted projection across 500 kpc (right). Each image has sides $300 \times 500 \rm kpc$. }
\label{fig:hydra_1}
\end{figure}

\begin{figure}
\includegraphics[width=0.45\textwidth,height=0.45\textwidth]{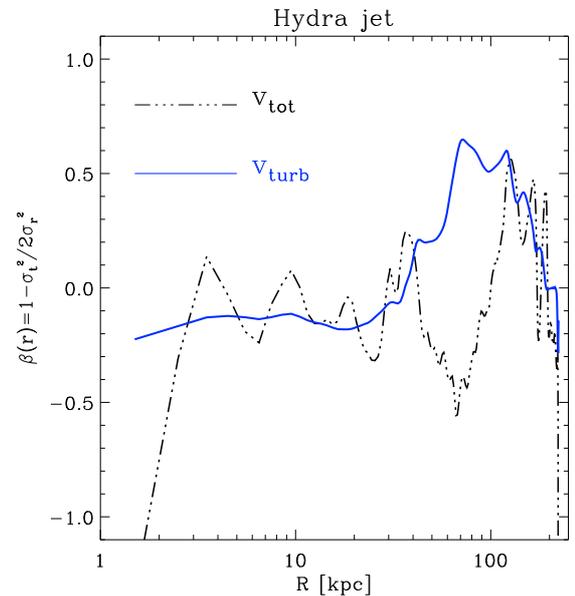}
\caption{Average radial profile of the anisotropy parameters of
turbulent motions in the Hydra A simulation. We show as a 
dashed line the profile of the total velocity field, and as a solid
line the profile of the turbulent velocity field.}
\label{fig:hydra_4}
\end{figure}
  
\subsection{AGN outflows in cluster cores}
\label{subsec:agn}

Powerful outflows from AGN are a viable mechanism to induce significant
turbulent motions in the innermost region of galaxy clusters, contributing
to the mixing of metals in the ICM, and to the lifting of cold and low-entropy materials to the outer cluster volume, thus reducing or quenching 
cluster cooling flows \citep[e.g.][]{co97,chu01,brh05,re06}. Only
recently fully cosmological grid simulations have reached a sufficient dynamical
range to model the evolution and feedback of AGN outflows in detail,
studying the interplay between AGNs and the ICM \citep[e.g.][]{xu09,tey11,dub11,kim11}. 

 Recently, the outflows from AGNs in the multiphase ICM has been simulated in detail with {\small FLASH} simulations by \citet{gaspari12a} and \cite{gaspari12b}. These authors reported typical values of $\sim 100-300 \rm ~km~s^{-1}$ for the turbulent velocity injected by the AGN, ruling the onset of non-linear instabilities and leading to the condensation of cold gas filaments.

\bigskip

We analysed the output of the simulated AGN-driven outflow in the Hydra A cluster, simulated
with {\small FLASH 3.2}. The volume around the jet injection has a side length of 1 Mpc and 
employs AMR to reach the maximum spatial resolution of 0.5 kpc per cell. The AGN jet is 
reconstructed by two circular back-to-back inflow boundaries 12 resolution
elements in diameter (2 kpc). The jet material is injected in opposite directions
(at a velocity of $v_{\rm jet}=3 \cdot 10^{4} \rm~km~s^{-1}$), with a total power of $W_{\rm jet}=3 \cdot 10^{45} \rm~erg~s^{-1}$.
At the epoch analysed here, the bulk velocity along the jet is $\sim 1500-1800\rm~km~s^{-1}$, and a powerful $M \sim 1.3$ shock has been
driven into the surrounding ICM.
To mimic the observed offset between the shock centre and the AGN, a smooth velocity field of $\sim 670\rm~km~s^{-1}$ has also been imposed
to the simulation, as a potential flow around a sphere of 100 kpc radius directed towards (-1,1,0). 
For more details of the simulation setup, we refer the reader to
\citet[][]{bruggen07} and \citet[][]{simi09}.

 In Fig.\ref{fig:hydra_1} we show maps of the module of the
 turbulent velocity field for a slice of depth 1 kpc (left) and for the volume-weighted projection along 500 kpc (right) for a region around the cluster centre.

 As in the Virgo runs, our method is very efficient in removing the
 large-scale ($>50$ kpc) laminar component of the velocity field, and highlights
 the complex pattern of turbulent structures associated to the interaction
 of the jet with the ICM atmosphere. Even if the  bulk velocity along the
 jet axis can be as high as $\sim 2000$ km/s, turbulence is on 
 average injected by  rolls of size $\sim 10-20$ kpc, via hydrodynamical
 instabilities, at the low velocity of $\sim 200-300 \rm~km~s^{-1}$. Weaker motions ($\sim 10-100\rm~km~s^{-1}$) are also injected in the innermost 100 kpc, perpendicular to the jet axis, following lateral expanding weak shock waves.  These findings are consistent with the recent ones of \citet{gaspari12a} and \citet{gaspari12b}.
 Some of the turbulent features of the outer jet structure
may be detected in nearby AGN inside the galaxy by {\small Astro-H} or {\small Athena} \citep[e.g.][]{heinz10,mendy11}. 
 
\subsubsection{Anisotropy of turbulence in AGN outflows}

Fig.\ref{fig:hydra_4} shows the average radial trend of $\beta(r)$ for the total (solid line)
and for the turbulent velocity field reconstructed with our filter (dot-dashed). The large-scale velocity field is dominated by radial
motions outside of 100 kpc (associated with the expansion of the jet and of the
running shock at the boundaries of the domain), with strong features 
of radial motions inside 100 kpc, connected to high-velocity "knots" along
the jet. The turbulent motions are close to isotropic along the whole jet structure, suggesting that instabilities are
very efficient in distributing the driving kinetic power of the jet in 3D.

This implies that the dissipation of kinetic energy in the surrounding ICM is very isotropic, which aids the AGN's capability of heating the cool core of the cluster.
However, the presence of an even weak magnetic field carried with the jet is expected to have a sizable effect, for instance affecting the exchange of
heat between the jet region and the surrounding ICM due to the local magnetic pressure gradient \citep[e.g.][]{oj10}.

\subsection{Power spectra of turbulence in AGN outflows}

\begin{figure}
\includegraphics[width=0.45\textwidth,height=0.45\textwidth]{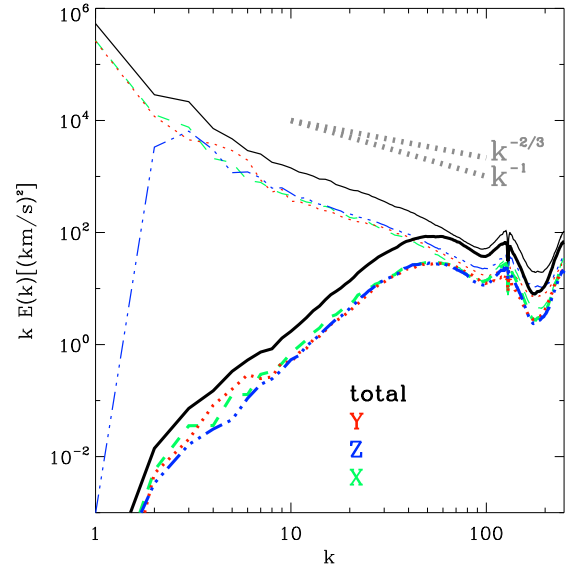}
\caption{Power spectra for the total velocity field (thin upper lines) and for
the turbulent velocity field (lower thick lines) for the same region
as in Fig.\ref{fig:hydra_1}. The different colour-coding shows the spectra for the component of velocities along the three axes of the simulation (direction 'Y' is the propagation axis of the jet).}
\label{fig:hydra_3}
\end{figure}

As before, we used the power spectrum of the turbulent velocity
field to investigate the important physical scales of the turbulence in the simulation volume (Fig.\ref{fig:hydra_3}).
We also decomposed the spectra
showing the contribution from the three velocity components separately.
While at very large scales there is some small excess of power
along the initial injection axis of the jet (Y),
as well as the lack of large-scale motions along the Z-axis (owing the large-scale velocity field along (-1,1,0) imposed by construction),
for most of the scales in the spectrum the power of 
the three components is very similar. 
Turbulence peaks at very small scales ($k \sim 50-100$, corresponding to $\sim 5-10$ kpc), but the resolution
of this run appears to be insufficient to measure the spectral shape of the turbulent cascade in a clear way.

\subsection{Turbulent diffusion for AGN outflows}

\begin{figure}
\includegraphics[width=0.45\textwidth]{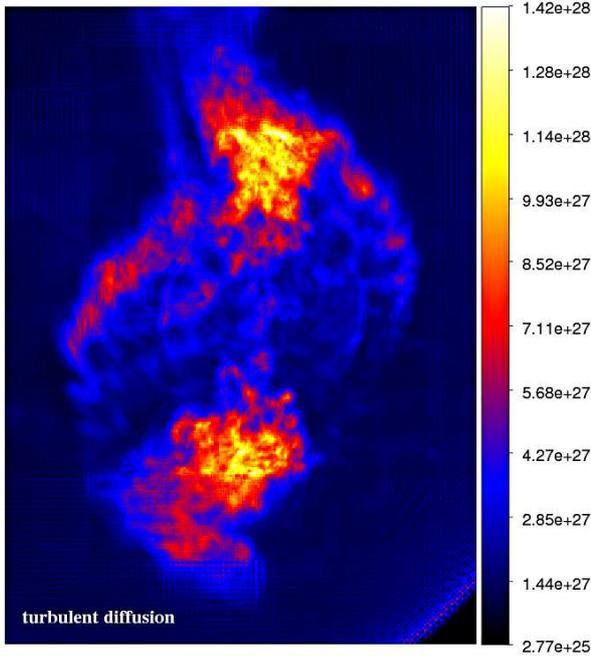}
\caption{Map of average turbulent diffusion (in [${\rm cm^{2}~s^{-1}}$])
for the projection along 500 kpc in the Hydra A run. The size of the image is the same as in Fig.\ref{fig:hydra_1}.}
\label{fig:hydra_2}
\end{figure}

Figure \ref{fig:hydra_2} shows the projected map of average turbulent diffusion across 500 kpc in the Hydra A run, measured as in the previous
sections. The highest values of turbulent diffusion on large scales are associated with the most prominent turbulent rolls in the range $\sim 50-100$ kpc from the centre, with $D_{\rm turb} \sim 1.5-2 \cdot 10^{28} {\rm cm^{2}~s^{-1}}$, while much lower values ($\sim 10^{27} {\rm cm^{2}~s^{-1}}$) are found in the inner 50-100 kpc around the
cluster core. 
Extended patterns of efficient diffusion are also associated to the 
motions perpendicular to the jet axis in the downstream region
of laterally expanding shock waves excited in the AGN outflows.
The values we measure for the Hydra A run are compatible with the 
turbulent diffusion necessary to explain the gradient of metallicity in the innermost cluster regions \citep[][]{re06,roe07}, provided
that what we measure here is turbulent diffusion $\sim 160$ Myr after the initial jet was launched in the simulation box.

\bigskip

\section{Conclusion}
\label{sec:conclusions}

We have presented and tested a simple and robust algorithm for
extracting the turbulent velocity field from a generic 3D
velocity field in grid simulations.
The algorithm is based on an iterative geometrical analysis of the velocity
field, and makes no a priori assumption on a physical
scale when filtering out laminar motions.
It iteratively calculates the local mean velocity and the
turbulent velocity for increasing volumes around each
cell of the simulation, until convergence is reached, taking
into account spurious steep gradients related to shock waves.

Our tests in Sec.\ref{subsec:test} show that our method performs well in reconstructing the morphology and 
spectral features of intermittent patterns of subsonic and transonic turbulent motions 
embedded in a large-scale background field. The main
limitations of the algorithm emerge in the presence of cuspy background velocity profiles (where the change of slope in the 
profile may be filtered as a small-scale turbulent fluctuations, if the variations occur on $\sim$ few cells), and if
the outer scale of turbulence is very close to the typical scale
of the laminar flow.

We applied our method to cosmological simulations of galaxy clusters with {\small ENZO 1.5} (Sect.\ref{subsec:enzo1}), to sloshing cool-core clusters (Sect.\ref{subsec:cool_core} and to AGN outflows (Sect.\ref{subsec:agn}) simulated with {\small FLASH 3.2}. 

In cosmological simulations, we find that turbulent velocities are slightly tangential in the inner regions and isotropic in regions close to the virial radius. The same is found for turbulence excited by cool-core sloshing, while a jet produces slightly radial turbulence and isotropic turbulence near its sonic point and beyond. 

\begin{figure*}
\includegraphics[width=0.95\textwidth]{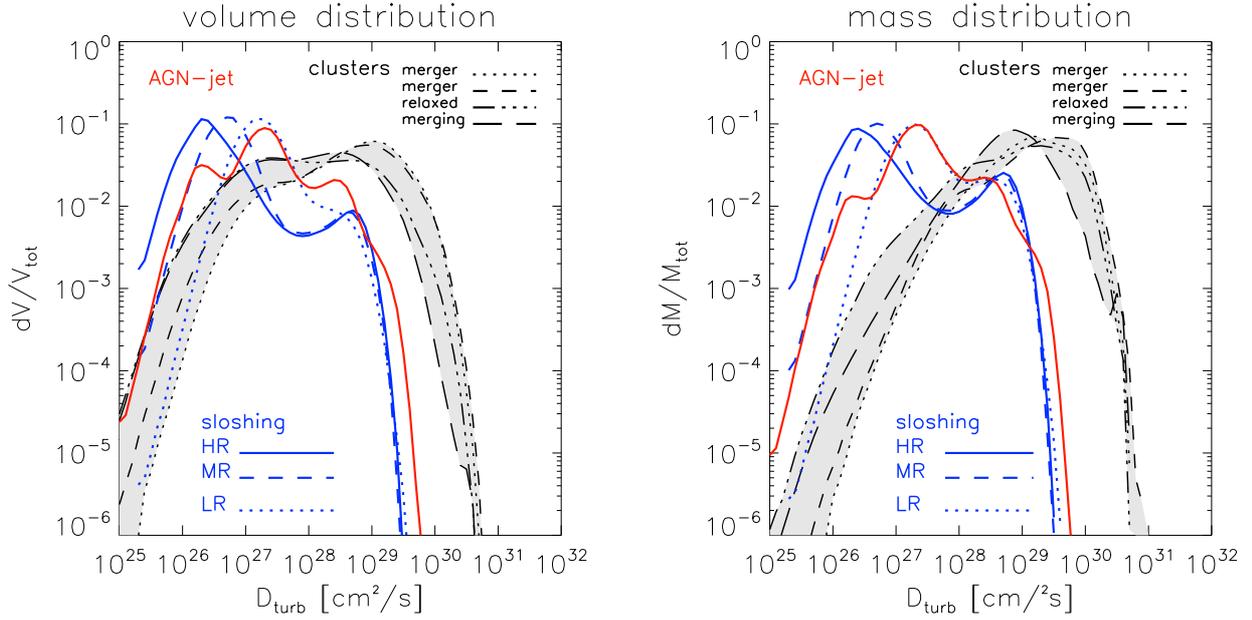}
\caption{Volume (left) and mass (right) distribution of turbulent diffusion for all runs studied in this paper. We show in red the turbulent diffusion from the
AGN-jet of Hydra, in blue the distributions of turbulent diffusion from the sloshing core in Virgo (we plot with different line-styles the distributions at different resolution, as in Sec.\ref{subsec:cool_core}), and in black the turbulent diffusion from cosmological clusters (the different line-styles are for each different object studied in Sec.\ref{subsec:enzo1}, while the shadowed region shows the uncertainty in the overall cluster sample).}
\label{fig:diffusion_total}
\end{figure*}

Our method naturally provides a way to estimate the turbulent diffusion in simulations, $D_{\rm turb} \propto \Lambda \cdot \sigma_{\rm v}$.
We show in Fig.\ref{fig:diffusion_total} the direct comparison of the
volume and mass distribution of turbulent diffusion in the three
cases. Each  mechanism presents a particular shape of the distribution. Turbulent diffusion from cluster mergers in general has a "simple" distribution with a maximum at $D_{\rm turb} \sim 10^{29} {\rm cm^{2}~s^{-1}}$ (volume-weighted distribution), with the tendency of post-merger systems to present tails of enhanced  diffusion,
up to several $\sim 10^{30} {\rm cm^{2}~s^{-1}}$.
In the  sloshing cool core, we observe two maxima in the distribution: one at $ \sim 10^{29} {\rm cm^{2}~s^{-1}}$ and associated with the innermost turbulent region
close to the cluster centre, and one with less efficient diffusion, $\sim 10^{26}-10^{27} {\rm cm^{2}~s^{-1}}$, associated with the KH rolls along the spiral
arms of the sloshing ICM. While the first feature is very stable against the change in resolution, the second one evolves with the increase of resolution because of the effect of a more efficient separation of differential rotation and turbulent KH rolls in our algorithm, and also because of the real onset of KH instabilities at
smaller scales in the simulation. 
The turbulent diffusion in the Hydra A jet is expected to be more
time-dependent compared to the other two. The distribution 
$\sim 160$ Myr after the jet launching presents a more complex distribution, owing to different patches of
fast diffusion in the jet-ICM regions of interactions.

Overall, the maximum values of turbulent diffusion attained by these mechanisms in the simulated ICM fall in the range
$D_{\rm turb} \sim 10^{29}-10^{30} {\rm cm^{2}~s^{-1}}$. Merger and accretion episodes in the ICM provide a more volume-filling mechanism of turbulent diffusion, and turbulent diffusion about one order of magnitude faster than jets and cool-core sloshing.
We note
that the average values we measure are of the order of the
upper limits derived with XMM-Newton observations of pseudo-pressure fluctuations in Coma ($D_{\rm turb} \leq 3 \cdot 10^{29} {\rm cm^{2}~s^{-1}}$, Schuecker et al. 2004), provided that our 3--D distributions of turbulent diffusion are usually patchy, and make a direct comparison non-trivial.  

A physical ingredient that can alter some of our findings
is the magnetic field. While its inclusion is not expected to change the dynamics of turbulent motions driven by large-scale mergers and accretion on $\gg ~{\rm kpc}$ scales \citep[e.g.][]{xu09,ruszkowski11,bonafede11}, local amplification of $\vec{B}$ in shear flows can suppress the growth of instabilities
and mixing motions along the spiral arms of sloshing structures
\citep[][]{zml11} and along AGN-jets \citep[][]{oj10}

\begin{figure}
\includegraphics[width=0.485\textwidth]{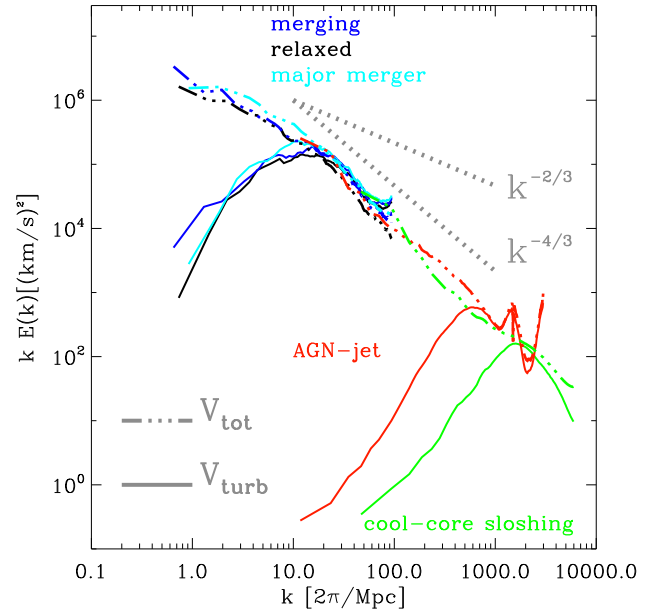}
\caption{Combined 3D power spectra from three clusters simulated with {\it {\small ENZO}} (Sec.\ref{subsec:enzo1}), for the {\it {\small FLASH}} simulations of the sloshing cool-core
in the Virgo cluster (Sec.\ref{subsec:cool_core}) and of the AGN-jet in the Hydra cluster (Sec.\ref{subsec:agn}). All spectra were normalised to have to obtain the same power at 200 kpc, while the wave numbers $k$ are the physical ones of each simulation.}
\label{fig:power_cosmic}
\end{figure}

We end by noting that to resolve the turbulence excited by cluster mergers, sloshing and AGN-jets in the same simulation and keeping the hydro-dynamic details presented in these runs, one would need to cover scales ranging from $R_{\rm vir} \sim 3$ Mpc down to the presumed scale of physical dissipation at $\sim 0.1$ kpc. As an
illustration of that, we show in Fig.~\ref{fig:power_cosmic} a composite power-spectrum of the simulated ICM, obtained by stacking the power spectra
of total velocity and of turbulence for the simulations analysed in this work. Since the masses of the clusters as well as the kinetic energy input differ, we rescaled the turbulent energy to the turbulent energy contained within $\sim (200 {\rm kpc})^3$ (which is well-captured in all runs) in the
  relaxed cluster H1.  This plot illustrates what might be the power spectrum
 for a cluster of total mass $\approx 3 \cdot 10^{14} M_{\odot}$ and radius $\approx 1900$ kpc, subject to sloshing event and a low-power jet of $W_{\rm jet} \sim 10^{44} {\rm ~erg~s^{-1}}$ in the last $\sim 100$ Myr (the power of the jet is estimated from the amount of the rescaling needed to match the spectrum of the Hydra run to that of the simulated cluster H1 at 200 kpc). 
 
  The obvious caveat is that because the different turbulent motions come from independent volumes, the interaction of turbulent
 modes along the turbulent cascade cannot be captured in this way. However, it is interesting that when rescaled in this
 way, the total velocity field from all simulations sits on a large-scale power-law, with $k \cdot E(k) \propto k^{-4/3}$ (equivalent to $E(k) \sim k^{-7/3}$), from $\sim 3-5$ Mpc to $\sim 0.5-1$ kpc. 
 Because in all simulations the background density profile is close
 to a beta-model, we propose that this broad-spectrum unveils the typical structure of correlated velocity field 
 imposed in the stratified ICM, mainly because of geometrical reasons. When turbulent motions are intermittently injected 
 into the ICM, they would add their spectral power to this  pre-existing power law, causing bumps in the spectrum.
They would also cause a flattening in the "background" velocity  spectrum along their turbulent cascade, producing a spectral slope close to
 $E(k) \sim k^{-5/3}-k^{-2}$.
 In the  near future, the efficient design of filtering methods to analyse this wide range of dynamical scales will be an important challenge for
the theoretical understanding of turbulent motions in cosmological simulations of large-scale structures.

\section{Acknowledgements}
F.V. and M.B. acknowledge support from the grant FOR1254 from the Deutsche Forschungsgemeinschaft (DFG).
E.R. acknowledges the support of the Priority Programme Witnesses of Cosmic History of the DFG, the supercomputing grants NIC 2877, 3229 and 3711 at the John-Neumann-Institut at the Forschungszentrum J\"ulich.
F.V. acknowledges using computational resources under the CINECA-INAF-2008/2010 agreement. Some of the results presented were produced using the FLASH code, a product of the DOE ASC/Alliances-funded Center for Astrophysical Thermonuclear Flashes at the University of Chicago. We acknowledge K. Dolag, I. Zhuravleva, J. Niemeyer, and W. Schmidt for fruitful technical discussions. We thank G. Brunetti, C. Gheller, and R. Brunino for their collaboration in producing the ENZO simulations analysed in this work.

\bibliography{scienzo}
\bibliographystyle{aa}

\bigskip

\bigskip

\appendix

{\bf Appendix}

Below we give an example of our multi-scale algorithm written in 
{\small IDL 7.0}, designed to recursively analyse one
velocity component of a 3--D field.  

The program takes as input a 3--D velocity field, {\tt vel} (we assume here it is written in binary format),  as a regular grid of linear
dimension of {\tt n} cells.  The fiducial tolerance parameters {\tt eps=0.1} and {\tt epssk=1} are used in the computation,
and a kernel scale of {\tt nk=8} is preliminarly used to compute the average skewness of the velocity field around each cell.

The final output of the code are 3--D distributions of the
turbulent field ({\tt turb}), of the outer scale ({\tt scale}), and of the skewness of the velocity field ({\tt sk}).

This code makes intensive use of the intrinsic IDL functions {\tt smooth}, {\tt convol}, and {\tt where} to reduce the usage of 3--D loops. We verified 
that this greatly speeds up the execution.

\begin{figure}
\includegraphics[width=0.4\textwidth]{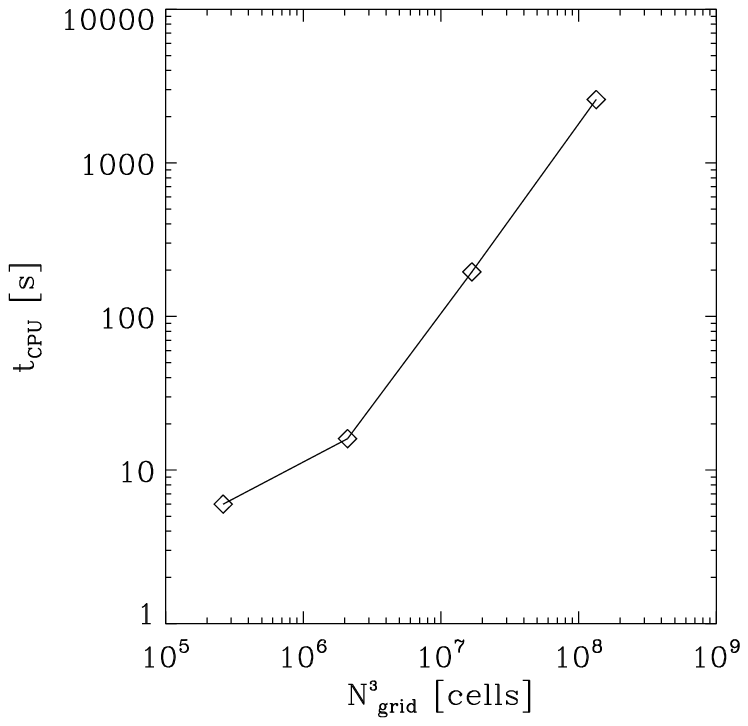}
\caption{Scaling between the CPU time employed by our algorithm and the
total number of cells in the computational volume for four different resolution of the Virgo HR run (Sec.\ref{subsec:cool_core}). Tests run on an Intel Quad-Core Xeon E5345 Linux Cluster.}
\label{fig:scal}
\end{figure}

In Fig.\ref{fig:scal} we show the benchmark tests
of the code applied to HR runs of the Virgo cluster (Sec.\ref{subsec:cool_core}) and interpolated to four different grid
resolutions of $64^{3}$, $128^{3}$, $256^{3}$ and $512^{3}$.  
For a sufficiently high resolution, the scaling of our algorithm
is very close to linear with respect to the number of cells analysed. 
The execution time to perform the multi-scale filter analysis of one component
of velocity in a $512^{3}$ grid is $\sim 45$ minutes, and $\sim 3$ minutes for a $256^{3}$ grid. 
 However, the details of the performance and scaling may change from problem to problem, depending on the
intermittency of the 3D velocity field under analysis.

The source code for our algorithm  can also be downloaded at this URL: 

http://www.ira.inaf.it/$\sim$vazza/papers/turbofilter.pro.

Also, a sample 3--D file of velocity extracted from a cluster simulation is given as an example,
at this URL: 
http://www.ira.inaf.it/$\sim$vazza/papers/sample\_velocity.dat.


\bigskip

{\tt
pro turbofilter

n=256  ;...linear size of the grid

vel=fltarr(n,n,n)  ;...3-D array of velocity

openr,3,'vel.dat' ;...the grid is read

readu,3,vel

close,3

;...needed parameters and thresholds

r2=uint(n*0.5-1.) ;...upper limit for L

r1=4           ;....lower limit 

turbo=fltarr(n,n,n)  ;....turbulent field

scale=fltarr(n,n,n) ;....scale of the flow

sk=fltarr(n,n,n)    ;....skewness

scale(*,*,*)= 0.    ;...initialization

turbo(*,*,*)=0.    ;...

eps=0.1        ;...tolerance in Eq.5
 
nk=8           ;...number of cells to compute skewness
 
epssk=1.        ;...tolerance for the skewness

drr=1.            ;...radial step for Eq.5

;... preliminary computation of the skewness

meanv = smooth(vel,nk,/EDGE\_TRUNCATE)
  
sc = abs((vx-meanv)/vel)
     
kernel=MAKE\_ARRAY(nk,nk,nk, /float, value = 1.)

kernel(0,*,*) = 0.

kernel(*,0,*) = 0.

kernel(*,*,0) = 0.
     
kernel(nk-1,*,*) = 0.
     
kernel(*,nk-1,*) = 0.
    
kernel(*,*nk-1) = 0.
     
sk=convol((vel-meanv)\textasciicircum2,kernel,/edge\_truncate)

sk=meanv\textasciicircum3/float(sk\textasciicircum1.5) ; skewness, Eq.7

sc1 = 0 ;.....

;....iterations to constrain turbulence

for r=r1,r2,drr do begin  

width = 2.*r+1 ; width of box
     
meanv = smooth(vel,width,/EDGE\_TRUNCATE) ;...mean local velocity at each scale
     
sc = abs((vel-meanv)/vel) ;...differential change in vel, Eq.5
     
skm=smooth(sk,width,/EDGE\_TRUNCATE) ;average skewness within L

;...check of which cells are converged

ibox=where((abs(sc-sc1)/float(sc1) lt eps or abs(skm) gt  epssk) and scalex eq 0.,nn)  
    
if nn gt 0 then begin
     
turbo(ibox)=vel(ibox)-meanv(ibox) ;...turbulent velocity in the cell
     
scale(ibox) = float(r+0.01)  ;...outer scale L

endif

sc1=sc

;..zc=n*0.5

;tvscl,[vel(*,*,zc),meanv(*,*,zc),turbo(*,*,zc)]

 endfor

;....saves our  final results

;...final turbulent field

  openw,3,"turb.dat"            

  writeu,3,turb

   close,3
 
;...outer scale

  openw,3,"scale.dat"         

  writeu,3,scale

   close,3

;...skewness 

  openw,3,"skewness.dat"     

  writeu,3,sk

   close,3

end

}

\bigskip
\end{document}